\documentclass[12pt,preprint]{emulateapj}

\usepackage{epsfig}
\usepackage{epstopdf}

\begin{document}

\def\beq{\begin{equation}}
\def\eeq{\end{equation}}

\newcommand{\pasa}{PASA}

\newcommand{\kms}{\,{\rm km\,s^{-1}}}
\newcommand{\msun}{\, M_\odot}
\newcommand{\lsun}{\, L_\odot}
\newcommand{\lvsun}{\, L_{\odot,V}}
\newcommand{\mlrsun}{\, M_{\odot} L_{\odot,R}^{-1}}
\newcommand{\mlvsun}{\, M_{\odot} L_{\odot,V}^{-1}}
\newcommand{\mbh}{M_\bullet}
\newcommand{\mstar}{M_\star}
\newcommand{\mbulge}{M_{\rm bulge}}
\newcommand{\ml}{M_\star/L}
\newcommand{\mlr}{M_\star/L_R}
\newcommand{\mlv}{M_\star/L_V}
\newcommand{\rinf}{\, r_{\rm inf}}
\newcommand{\reff}{\,r_{\rm eff}}
\newcommand{\scut}{\sigma_{\rm cut}}
\newcommand {\gtsim} {\ {\raise-.5ex\hbox{$\buildrel>\over\sim$}}\ }
\newcommand {\ltsim} {\ {\raise-.5ex\hbox{$\buildrel<\over\sim$}}\ }

\title{Revisiting the Scaling Relations of Black Hole Masses and Host Galaxy Properties }

\author{Nicholas J. McConnell \footnotemark[1,2] and Chung-Pei Ma \footnotemark[2]}

\footnotetext[1]{Institute for Astronomy, University of Hawaii at Manoa, Honolulu, HI; nmcc@ifa.hawaii.edu}
\footnotetext[2]{Department of Astronomy, University of California at Berkeley, Berkeley, CA; cpma@berkeley.edu}

\begin{abstract}

 New kinematic data and modeling efforts in the past few years have
  substantially expanded and revised dynamical measurements of black hole
  masses ($\mbh$) at the centers of nearby galaxies.  Here we compile an
  updated sample of 72 black holes and their host galaxies, and present
  revised scaling relations between $\mbh$ and stellar velocity dispersion
  ($\sigma$), $V$-band luminosity ($L$), and bulge stellar mass
  ($\mbulge$), for different galaxy subsamples.  Our best-fitting power law
  relations for the full galaxy sample are $\log_{10}(\mbh) = 8.32 + 5.64
  \log_{10}(\sigma/200 \kms)$, $\log_{10}(\mbh) = 9.23 + 1.11
  \log_{10}(L/10^{11}\lsun)$, and $\log_{10}(\mbh) = 8.46 + 1.05
  \log_{10}(\mbulge/10^{11}\msun)$.  A log-quadratic fit to the
  $\mbh-\sigma$ relation with an additional term of $\beta_2 \,
  [\log_{10}(\sigma/200 \kms)]^2$ gives $\beta_2=1.68 \pm 1.82$ and does not
  decrease the intrinsic scatter in $\mbh$.  
  Including 92 additional upper limits on $\mbh$ does not change the slope of the $\mbh-\sigma$ relation.  
  When the early- and late-type galaxies are fit
  separately, we obtain similar slopes of 5.20 and 5.06 for the
  $\mbh-\sigma$ relation but significantly different intercepts --
  $\mbh$ in early-type galaxies are about 2 times higher than in late types at a given sigma.
  Within early-type galaxies, our fits to $\mbh(\sigma)$ give $\mbh$ that is about 2 times higher in galaxies with central core profiles than those with central power-law profiles.
  Our $\mbh-L$ and $\mbh-\mbulge$ relations for early-type galaxies are similar to those from earlier compilations, and core and power-law galaxies yield similar $L$- and $\mbulge$-based predictions for $\mbh$.    
When the conventional quadrature method is used to determine the intrinsic scatter in $\mbh$, our dataset 
shows weak evidence for increased scatter at $\mbulge < 10^{11} \msun$ or $L_V < 10^{10.3} \lsun$, while
the scatter stays constant for $10^{11} <  \mbulge < 10^{12.3} \msun$
and $10^{10.3} < L_V < 10^{11.5} \lsun$.   A Bayesian analysis indicates that a larger sample of $\mbh$ measurements 
would be needed to detect any statistically significant trend in the scatter with galaxy properties.

\end{abstract}

\pagestyle{plain}
\pagenumbering{arabic}

\maketitle

\section{Introduction}
\label{sec:intro}

Empirical correlations between the masses, $\mbh$, of supermassive black
holes and different properties of their host galaxies have proliferated in
the past decade.  Power-law fits to these correlations
provide efficient means to estimate $\mbh$ in large samples of galaxies, or in individual objects with insufficient data to measure $\mbh$ from the dynamics of stars, gas, or masers.  

Correlations between black hole masses and numerous properties of their host galaxies have been explored in the literature.
These include scaling relations between $\mbh$ and  
stellar velocity dispersion
(e.g., \citealt{Ferr00,Geb00,MF01,Tremaine02,Wyithe06a,Wyithe06b,Hu08,Gultekin}, hereafter G09; \citealt{Schulze11,mcconnell11b,Graham11, Beifiori12}, hereafter B12)
and between $\mbh$ and the stellar mass of the bulge
(e.g., \citealt{Magorrian,MH03,HRix,Hu09,Sani11}; B12).
%
Various scaling relations between $\mbh$ and
the photometric properties of the galaxy have also 
been examined:  bulge optical luminosity 
(e.g., \citealt{KR95,KG01}; G09; \citealt{Schulze11,mcconnell11b}; B12),
%
bulge near-infrared luminosity
\citep[e.g.,][]{MH03,MD02,MD04,Graham07,Hu09,Sani11},
total luminosity 
(e.g., \citealt{KG01,KBC11}; B12),
%
and bulge concentration or S\'{e}rsic index
(e.g., \citealt{Graham01,GD07}; B12).
%
On a larger scale, correlations between $\mbh$ and the circular velocity or
dynamical mass of the dark matter halo have been reported as well as disputed
(e.g., \citealt{Ferr02,Baes03,Zasov05,KB11,Volonteri11}; B12).
%
More recently, $\mbh$ has been found to correlate with the number and total mass of
globular clusters in the host galaxy \citep[e.g.,][]{BT10,HH11,SC12}.
In early-type galaxies with core profiles, \citet{Lauer07} and \citet{KB09} have explored
correlations between $\mbh$ and the core radius, or the total ``light
deficit'' of the core relative to a S\'{e}rsic profile.

Recent kinematic data and modeling efforts have substantially 
expanded the various samples used in all of the studies above.
In this paper, we take advantage of these developments, presenting
an updated compilation of 
72  
black holes and their host galaxies and providing new scaling relations.
Our sample is a significant update from two recent
compilations by G09 and \citet{Graham11}.
Compared with the 49 objects in
G09, 27 black holes in our present sample are new measurements, and 18
masses are updated values from better data and/or more sophisticated
modeling.  Compared with the 64 objects in \citet{Graham11} (an update of
\citealt{Graham08}), 35 of our black hole masses are new or updates.  
The G09 and \citet{Graham08} samples
differ by only a few galaxies, based on the authors' respective judgments
about which dynamical measurements are reliable.
The most significant updates in our sample are galaxies with extremely
high $\mbh$ \citep{SG10,Geb11,mcconnell11a,mcconnell11b, mcconnell12,RusliThesis} and
galaxies with some of the smallest observed central black holes
\citep{Greene10,Nowak10,KBC11,Kuo11}.  
Our present sample includes updated distances to 44 galaxies, mostly based on surface brightness fluctuation measurements \citep{Tonry01,Blakeslee09,Blakeslee10}.

We focus on three frequently studied scaling relations: $\mbh$
vs. stellar velocity dispersion ($\sigma$), $V$-band bulge luminosity
($L$), and stellar bulge mass ($\mbulge$).  As reported below, our new
compilation results in a significantly steeper power law for the
$\mbh-\sigma$ relation than in G09 and the recent investigation by Beifiori et al. (2012; B12), who combined the
previous sample of 49 black holes from G09 with a larger sample of upper
limits on $\mbh$ from \citet{Beifiori09}.  We still find a steeper power
law than G09 or B12 when we include these upper limits in our fit to the $\mbh-\sigma$
relation.  
We have performed a quadratic fit to $\mbh(\sigma)$ and find a marginal amount of upward curvature, similar to previous investigations (Wyithe 2006a,b; G09).

Another important measurable quantity is the intrinsic or cosmic scatter in
$\mbh$ for fixed galaxy properties.  Quantifying the scatter in $\mbh$ is
useful for identifying the tightest correlations from which to predict
$\mbh$ and for testing different scenarios of galaxy and black hole growth.
In particular, models of stochastic black hole and galaxy growth via
hierarchical merging predict decreasing scatter in $\mbh$ as galaxy mass
increases \citep[e.g.,][]{Peng07,JM11}.  Previous empirical studies of the
black hole scaling relations have estimated the intrinsic scatter in $\mbh$
as a single value for the entire sample.  Herein, we take advantage of our
larger sample to estimate the scatter as a function of $\sigma$, $L$, and
$\mbulge$.

In \S2 we summarize our updated compilation of 72 black hole mass
measurements and 35 bulge masses from dynamical studies.  In \S3 we present
fits to the $\mbh-\sigma$, $\mbh-L$, and $\mbh-\mbulge$ relations and
highlight subsamples that yield interesting variations in the best-fit
power laws.  In particular, we examine different cuts in $\sigma$, $L$, and
$\mbulge$, as well as cuts based on galaxies' morphologies and surface
brightness profiles.  In \S4 we discuss the scatter in $\mbh$ and its
dependence on $\sigma$, $L$, and $\mbulge$.  In \S5 we discuss how our
analysis of galaxy subsamples may be beneficial for various applications of
the black hole scaling relations.

Our full sample of black hole masses and galaxy properties is available
online at {\tt http://blackhole.berkeley.edu}.  This database will be
updated as new results are published.  Investigators are encouraged to use
this online database and inform us of updates.

\vspace{0.1in}

\section{An Updated Black Hole and Galaxy Sample}
\label{sec:newsample}

%
%
\begin{figure*}
 \vspace{-0.5in}
 \hspace{-0.3in}
  \epsfig{figure=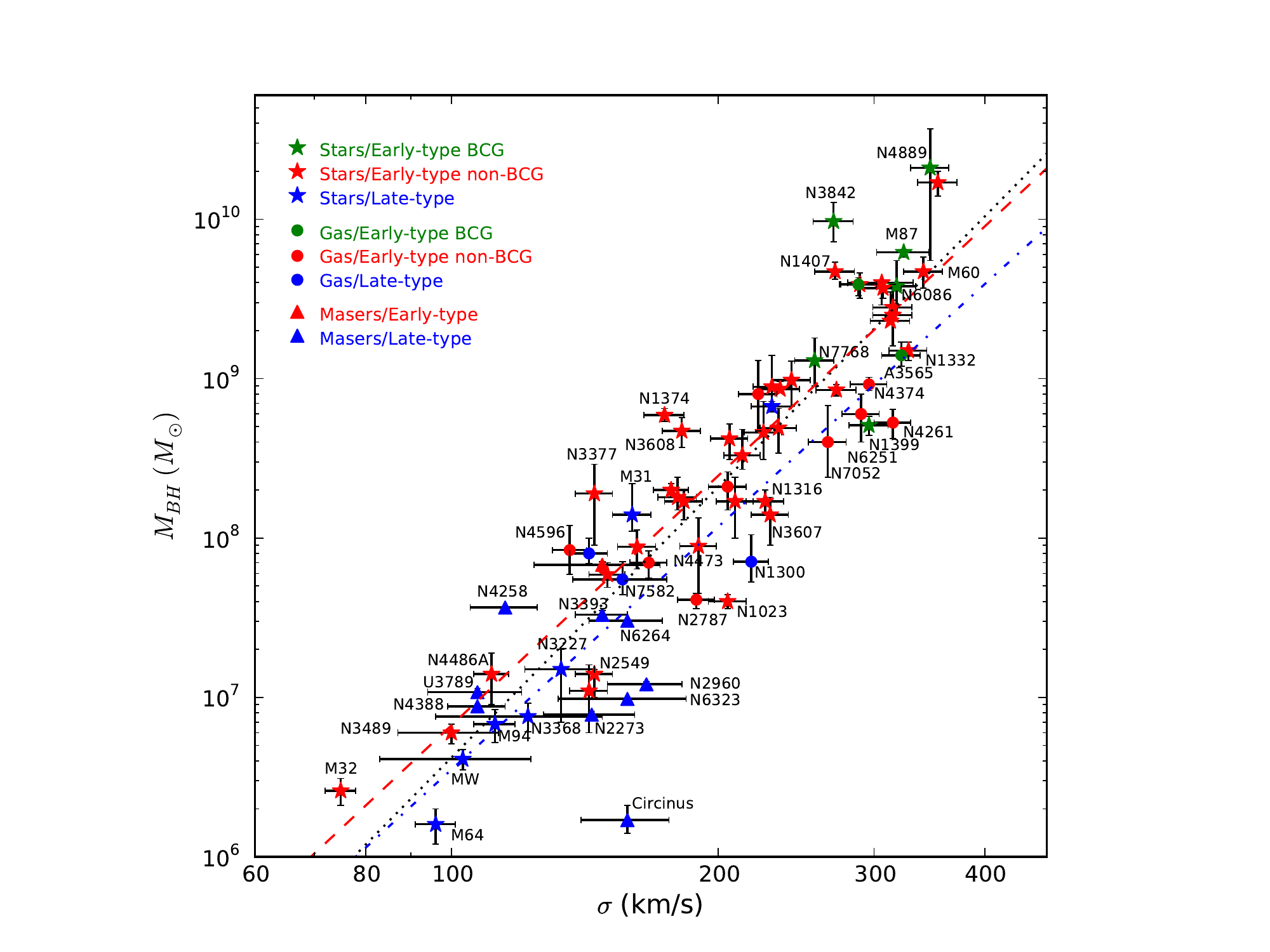,width=8.2in}

 \caption{The $\mbh-\sigma$ relation for our full sample of 72 galaxies listed in Table~\ref{tab:sample} and at {\tt http://blackhole.berkeley.edu}.  Brightest cluster galaxies (BCGs) that are also the central galaxies of their clusters are plotted in green, other elliptical and S0 galaxies are plotted in red, and late-type spiral galaxies are plotted in blue.   
NGC 1316 is the most luminous galaxy in the Fornax cluster, but it lies at the cluster outskirts;
the green symbol here labels the central galaxy NGC 1399.  
M87 lies near the center of the Virgo cluster, whereas NGC 4472 (M49) lies $\sim 1$ Mpc to the south.
The black-hole masses are measured using the dynamics of masers
(triangles), stars (stars) or gas (circles). Error bars indicate 68\%
confidence intervals. For most of the maser galaxies, the error bars in
$\mbh$ are smaller than the plotted symbol.  The black dotted line shows the
best-fitting power law for the entire sample: $\log_{10}(\mbh/\msun) = 8.32
+ 5.64\log_{10} (\sigma/200 \kms)$. When early-type and late-type galaxies
are fit separately, the resulting power laws are $\log_{10}(\mbh/\msun)
= 8.39 + 5.20\log_{10} (\sigma/ 200 \kms)$ for the early-type (red dashed line),
and $\log_{10}(\mbh/\msun) = 8.07 + 5.06\log_{10} (\sigma/ 200 \kms)$ for
the late-type (blue dot-dashed line).  The plotted values of $\sigma$ are derived
using kinematic data over the radii $\rinf < r < \reff$.}
\label{fig:plotmsig}
\vspace{0.1in}
\end{figure*}

\begin{figure*}
 \vspace{-0.5in}
  \hspace{-0.3in}
  \epsfig{figure=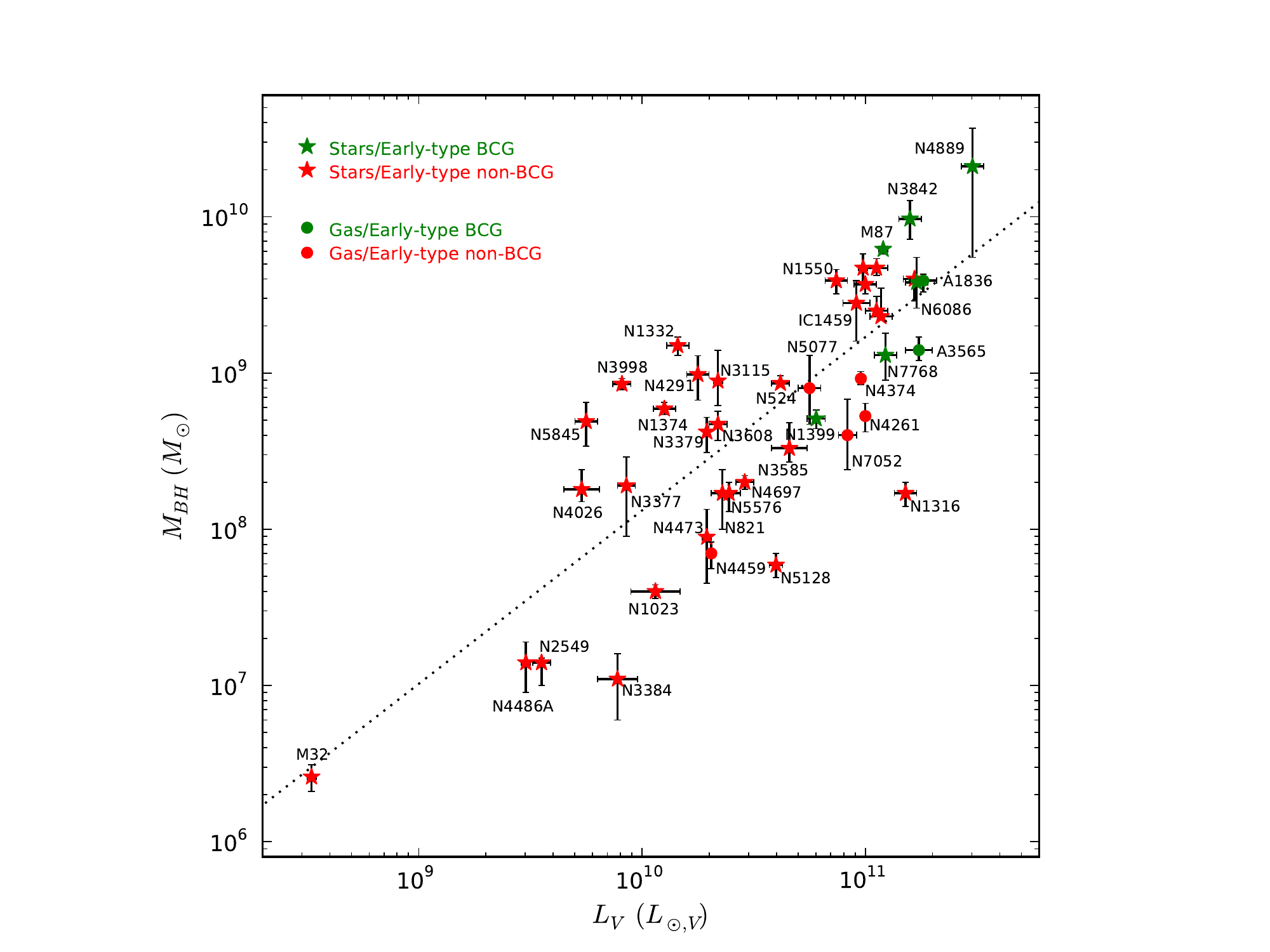,width=8.2in}
\caption{The $\mbh-L$ relation for the 44 early-type galaxies with reliable
  measurements of the V-band bulge luminosity in our sample.  The symbols
  are the same as in Figure 1.  The black line represents the best-fitting
  power-law $\log_{10}(\mbh/\msun) = 9.23 + 1.11\,\log_{10} (L_v/10^{11}
  \lsun)$.  }
 \label{fig:plotml}
\vspace{0.1in}
\end{figure*}

\begin{figure*}
  \vspace{-0.5in}
 \hspace{-0.3in}
  \epsfig{figure=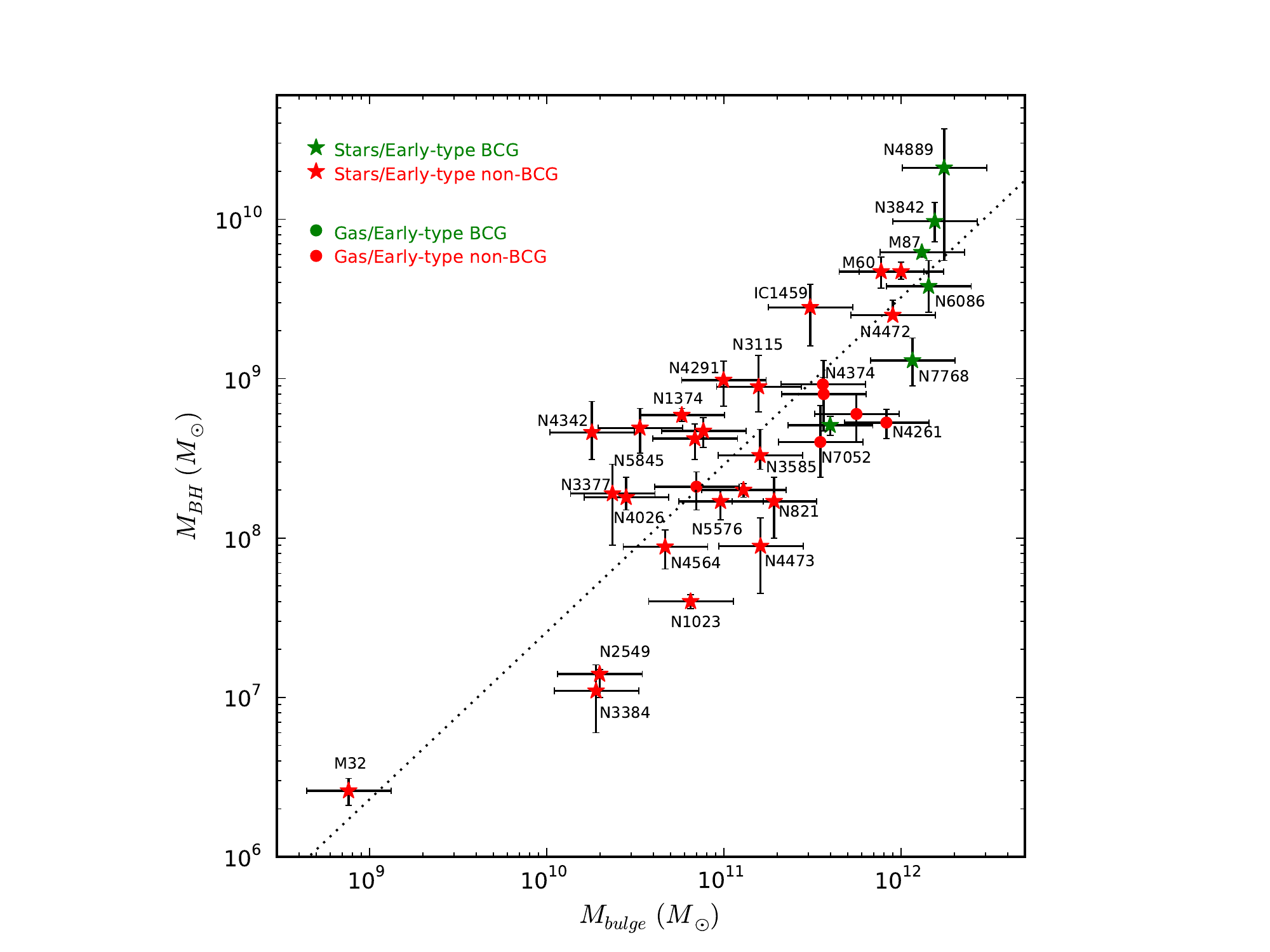,width=8.2in}
\caption{The $\mbh-\mbulge$ relation for the 35 early-type galaxies with
  dynamical measurements of the bulge stellar mass in our sample.  The
  symbols are the same as in Figure 1.  The black line represents the
  best-fitting power-law $\log_{10}(\mbh/\msun) = 8.46 + 1.05\,\log_{10}
  (\mbulge/10^{11} \msun)$.  }
 \label{fig:plotmm}
\vspace{0.2in}
\end{figure*}

Our full sample of 72 black hole masses and their host galaxy properties are listed in
Table~\ref{tab:sample}, which appears at the end of this paper.  The corresponding $\mbh$ versus $\sigma$, $L$, and
$\mbulge$ are plotted in Figures~\ref{fig:plotmsig}-\ref{fig:plotmm}.  This
sample is an update of our previous compilation of 67 dynamical black hole
measurements, presented in the supplementary materials to
\citet{mcconnell11b}.  The current sample includes one new measurement of
$\mbh$ from \citet{mcconnell12}, seven new measurements from \citet{RusliThesis}, and two updated measurements (NGC
4594, \citealt{Jardel11}; NGC 3998, \citealt{walsh12}).  For NGC 5128 (Cen
A), we have adopted the value $\mbh = 5.9^{+1.1}_{-1.0} \times 10^7 \msun$
(at a distance of 4.1 Mpc) from \citet{Capp09}.

We have removed three galaxies whose original measurements have exceptional
complications.  \citet{LB03} measured non-Keplerian maser velocities in NGC
1068 and estimated $\mbh$ by modeling a self-gravitating disk.  Still,
other physical processes might reproduce the observed maser motions.
\citet{atkinson05} reported a measurement of $\mbh$ in NGC 2748 but noted
that heavy extinction could corrupt their attempt to locate the center of
the nuclear gas disk.  \citet{Geb03} justified classifying the central
point source of NGC 7457 as an active galactic nucleus, but their arguments
permit the central mass to be shared by an accreting black hole and a
nuclear star cluster.  

Additionally, we have updated the distances to 44 galaxies in our sample.
For 41 galaxies, we adopt surface brightness fluctuation measurements from
\citet{Tonry01} and \citet{Blakeslee09}, with the corrections suggested by
\citet{Blakeslee10}.  For M31 and M32, we adopt the Cepheid variable
distance of 0.73 Mpc from \citet{Vilardell07}.  For NGC 4342, we adopt the
distance of 23 Mpc from \citet{Bogdan12b}.  Other measured quantities are
scaled accordingly: $\mbh \propto D$, $L \propto D^2$, and $\mbulge \propto
D$.  Table~\ref{tab:sample} includes the updated values for all quantities.
The new galaxy distances and rescaled $\mbh$ only have a small effect on
our fits to the black hole scaling relations.  For other galaxy distances, we assume $H_0 = 70 \kms$ Mpc$^{-1}$, as in \citet{mcconnell11b}.

For the $\mbh-\sigma$ relation, we also consider upper limits for $\mbh$ in
89 galaxies from B12, plus 3 new upper limits
\citep{Schulze11,Gultekin11b,mcconnell12}.  Five additional galaxies in the
B12 upper limit sample have recently obtained secure
measurements of $\mbh$ and are included in our 72-galaxy sample.
As we discuss in \S3, including upper limits results in a lower normalization (intercept) for the $\mbh-\sigma$ relation but does not significantly alter the slope.

%

For the $\mbh-\sigma$ relation, we consider two different definitions of
$\sigma$.  Both definitions use spatially resolved measurements of the
line-of-sight velocity dispersion $\sigma(r)$ and radial velocity $v(r)$,
integrated out to one effective radius ($\reff$):
 \beq \sigma^2 \equiv \frac{\int^{\reff}_{r_{\rm
      min}} \left[\sigma^2(r) + v^2(r) \right]I(r)dr}{\int^{\reff}_{r_{\rm
      min}} I(r)dr} \;\; ,
\label{eq:sigeff}
\eeq
where $I(r)$ is the galaxy's one-dimensional stellar surface brightness
profile.  In G09 and most other studies, the lower integration limit
$r_{\rm min}$ is set to zero and sampled at the smallest scale allowed by
the data.  This definition of $\sigma$, however, includes signal from
within the black hole radius of influence, $\rinf \equiv G\mbh\sigma^{-2}$.
In some galaxies, particularly the most massive ellipticals, $\sigma$
decreases substantially when spatially resolved data within $\rinf$ are
excluded. Setting $r_{\rm min} = \rinf$ produces an alternative definition
of $\sigma$ that reflects the global structure of the galaxy and is
less sensitive to angular resolution.
We compare the two definitions of $\sigma$ for 12 galaxies whose
kinematics within $\rinf$ are notably different from kinematics at larger
radii.  As shown in Table~\ref{tab:rinf}, excluding $r < \rinf$ can reduce
$\sigma$ by up to 10-15\%.
Ten of the 12 updated galaxies are massive ($\sigma > 250 \kms$ using either
definition).  \citet{RusliThesis} presented seven new stellar dynamical measurements of $\mbh$ along with central velocity dispersions.  We have used the long-slit kinematics from \citet{RusliThesis} and references therein to derive $\sigma$ according to Equation~\ref{eq:sigeff}; our $\sigma$ values appear in Tables~\ref{tab:sample} and~\ref{tab:rinf}.

For the $\mbh-\mbulge$ relation, we have compiled the bulge stellar masses
for 35 early-type galaxies.  Among them, 13 bulge masses are taken from \citet{HRix},
who used spherical Jeans models to fit stellar kinematics.  For 22 more
galaxies, we multiply the $V$-band luminosity in Table~\ref{tab:sample}
with the bulge mass-to-light ratio ($M/L$) derived from kinematics and
dynamical modeling of stars or gas (see Table~\ref{tab:sample} for
references).  Where necessary, $M/L$ is converted to $V$-band using galaxy
colors.  The values of $\mbulge$ are scaled to reflect the assumed
distances in Table~\ref{tab:sample}.

Most of the dynamical models behind our compiled values of $\mbulge$ have
assumed that mass follows light.  This assumption can be appropriate in the
inner regions of galaxies, where dark matter does not contribute
significantly to the total enclosed mass.  Still, several measurements are
based on kinematic data out to large radii.  Furthermore, some galaxies
exhibit contradictions between the dynamical estimates of $M/L$ and
estimates of $M/L$ from stellar population synthesis models
\citep[e.g.,][]{Capp06,CvD12}.  For this reason, we adopt a conservative
approach and assign a minimum error of 0.24 dex to each value of $\mbulge$.
The corresponding confidence interval (0.58 - 1.74) $\times \, \mbulge$ spans a factor of 3.

To test how well our $\mbulge$ values represent
the stellar mass of each galaxy, we also have fit the $\mbh-\mbulge$ relation
using a sample of 18 galaxies for which $\mbulge$ is computed from the
stellar mass-to-light ratio, $M_\star/L$.  
Our stellar $\mbulge$ sample comprises 13 galaxies for which $M_\star/L$ is measured from dynamical models including dark matter, plus five galaxies for which $M_\star/L$ is derived from stellar population models by \citet{CvD12}.  
This sample yields a slightly steeper slope of $1.34 \pm 0.15$ for the $\mbh-\mbulge$ relation, versus a slope of $1.05 \pm 0.11$ for our 35-galaxy dynamical $\mbulge$ sample.  The stellar $\mbulge$ sample also has substantially lower scatter in $\mbh$ (see Table~\ref{tab:fits}).

%
\begin{table}[!b]
\begin{center}
\caption{Galaxies with multiple definitions of $\sigma$}
\label{tab:rinf}
\begin{tabular}[b]{lclcc}  
\hline
\bf{Galaxy} & \bf{Ref.} & \bf{$\rinf$} & \bf{$\sigma$ (0-$\reff$)} & \bf{$\sigma$ ($\rinf$-$\reff$)}  \\ 
& & ($''$) & ($\kms$) & ($\kms$) \\
\hline 
\\
IC 1459 & 1 & 0.81 & 340 & 315 \\
NGC 1374 & 2 & 0.89 & 203 & 174 \\
NGC 1399 & 3,4 & 0.63 & 337 & 296 \\
NGC 1407 & 5 & 1.9 & 283 & 274 \\
NGC 1550 & 6 & 0.78 & 302 & 289 \\
NGC 3842 & 7 & 1.2 & 275 & 270 \\
NGC 3998 & 8 & 0.71 & 286 & 272 \\
NGC 4486 & 9 & 2.1 & 375 & 324 \\
NGC 4594 & 10 & 1.2 & 240 & 230 \\
NGC 4649 & 11 & 2.2 & 385 & 341 \\
NGC 4889 & 7 & 1.5 & 360 & 347 \\
NGC 7619 & 12 & 0.39 & 324 & 313 \\
NGC 7768 & 7 & 0.14 & 265 & 257 \\
\\
\hline   
\end{tabular}
\end{center}
Notes: References for kinematic data used to derive $\rinf$ are (1)
Cappellari et al. 2002; (2) D'Onofrio et al. 1995; (3) Graham et al. 1998; (4) Gebhardt et al. 2007;
(5) Spolaor et al. 2008; (6) Simien \& Prugniel 2000; (7) McConnell et al. 2012; (8) Walsh et al. 2012; (9) Gebhardt et al. 2011;
(10) Jardel et al. 2011; (11) Pinkney et al. 2003; (12) Pu et al. 2010.
Although \citet{RusliThesis} report long-slit kinematic measurements for NGC 1374, the measurements from \citet{Donofrio95} are more consistent with high-resolution SINFONI data in \citet{RusliThesis}.
\end{table}

\vspace{0.1in}
\section{Black Hole Scaling Relations and Fits}
\label{sec:fits}

In this section we present results for the fits to black hole
scaling relations for
the full sample of dynamically measured $\mbh$
listed in Table~\ref{tab:sample},
the full sample of $\mbh$ plus 92 upper limits on $\mbh$,
and various subsamples divided by galaxy properties.

\subsection{Fitting methods}
\label{sec:methods}

Our power law fit to a given sample is defined in log space by an intercept
$\alpha$ and slope $\beta$:
\beq
 \log_{10}  \mbh = \alpha + \beta \, \log_{10} X \,,
\label{eq:msigma}
\eeq where $\mbh$ is in units of $\msun$, and $X=\sigma/200\kms$,
$L/10^{11} L_\odot$, or $\mbulge/10^{11} M_\odot$ for the three scaling
relations.  We have also tested a log-quadratic fit for the $\mbh-\sigma$
relation:
\begin{eqnarray}
 \log_{10}  \mbh  =  \alpha + \beta \, \log_{10} X + 
            \beta_2 \, \left[ \log_{10} X \right]^2 \,, 
\label{eq:quadratic}
\end{eqnarray}
where $X=\sigma/200\kms$.  Results for the quadratic fit are discussed separately in
Sec.~\ref{sec:quadratic} below.

For the power-law scaling relations, we have compared three linear
regression estimators: {\tt MPFITEXY}, {\tt LINMIX\_ERR}, and {\tt BIVAR
  EM}.  {\tt MPFITEXY} is a least-squares estimator by \citet{Williams10}.
{\tt LINMIX\_ERR} is a Bayesian estimator by \citet{Kelly07}.  Both {\tt
  MPFITEXY} and {\tt LINMIX\_ERR} consider measurement errors in two
variables and include an intrinsic scatter term, $\epsilon_0$, in
log($\mbh$).  {\tt LINMIX\_ERR} can be applied to galaxy samples with upper
limits for $\mbh$.  For the $\mbh-\sigma$ sample with upper limits, we also
use the {\tt BIVAR EM} algorithm in the {\tt ASURV} software package by
\citet{lavalley92}, which implements the methods presented in
\citet{Isobe86}.  The {\tt ASURV} procedures do not consider measurement
errors, and we use this method primarily for comparison with B12.  All
three algorithms are publicly available\footnote{The IDL source code for
  {\tt MPFITEXY} is available at {\tt http://purl.org/mike/mpfitexy} .  THE
  IDL source code for {\tt LINMIX\_ERR} and dependent scripts is available
  at {\tt http://idlastro.gsfc.nasa.gov/ftp/pro/math} .  {\tt ASURV} is
  available at {\tt http://www2.astro.psu.edu/statcodes/asurv} .}.

For each of the global scaling relations and galaxy subsamples, we obtain
consistent fits from {\tt MPFITEXY} and {\tt LINMIX\_ERR}, although {\tt LINMIX\_ERR} usually returns a slightly higher value of $\epsilon_0$.
Table~\ref{tab:fits} includes the global fitting results from both methods.  In Table~\ref{tab:fits} we also include results from {\tt LINMIX\_ERR} in cases where $\epsilon_0$ is poorly constrained by {\tt MPFITEXY}.
For the $\mbh-\sigma$ relation including upper limits, the {\tt BIVAR EM} procedure returns a lower intercept than {\tt LINMIX\_ERR}, but the slopes from the two methods are consistent within errors.   
Recently, \citet{Park12} investigated the $\mbh-\sigma$ relation using four linear regression
estimators, including {\tt MPFITEXY} and {\tt LINMIX\_ERR}.  All four
estimators yielded consistent fits to empirical data, and {\tt MPFITEXY}
and {\tt LINMIX\_ERR} behaved robustly for simulated data with large
measurement errors in $\sigma$.

\subsection{$\mbh - \sigma$ Relation}
\label{sec:msigma}

Our fits to $\mbh(\sigma)$ for the entire galaxy sample and various
subsamples are plotted in Figures~\ref{fig:plotmsig} and~\ref{fig:corepl}a,
and summarized in Table~\ref{tab:fits}.  

\subsubsection{Full Sample}

Our full sample of 72 galaxies yields an intercept $\alpha = 8.32 \pm 0.05$
and slope $\beta = 5.64 \pm 0.32$.  
When upper limits are added, the sample of 164 galaxies yields $\alpha = 8.15 \pm 0.05$ and $\beta = 5.58 \pm 0.30$.
The reduced intercept is a natural consequence of considering upper limits, while the slightly shallower slope is consistent within errors.

\subsubsection{Early vs Late Type}

Fitting early- and late-type galaxies separately yields slightly shallower
slopes: $\beta = 5.20 \pm 0.36$ for early-types (red dashed line in Figure~\ref{fig:plotmsig})
and $\beta = 5.06 \pm 1.16$ for late-types (blue dot-dashed line).
The late-type galaxies have a significantly lower intercept: $\alpha=8.39 \pm 0.06$ versus $8.07 \pm 0.21$.  
Correspondingly, our fits predict $M_{\bullet, \rm early}
\sim 2 \, M_{\bullet, \rm late}$ at fixed $\sigma$.  Because most of the
late-type bulges have low $\sigma$, the split in intercepts leads to a
steeper slope of 5.64 for the full sample.

\subsubsection{Core vs Power-law}

\begin{figure}[!b]
\centering
\vspace{0.2in}
  \epsfig{figure=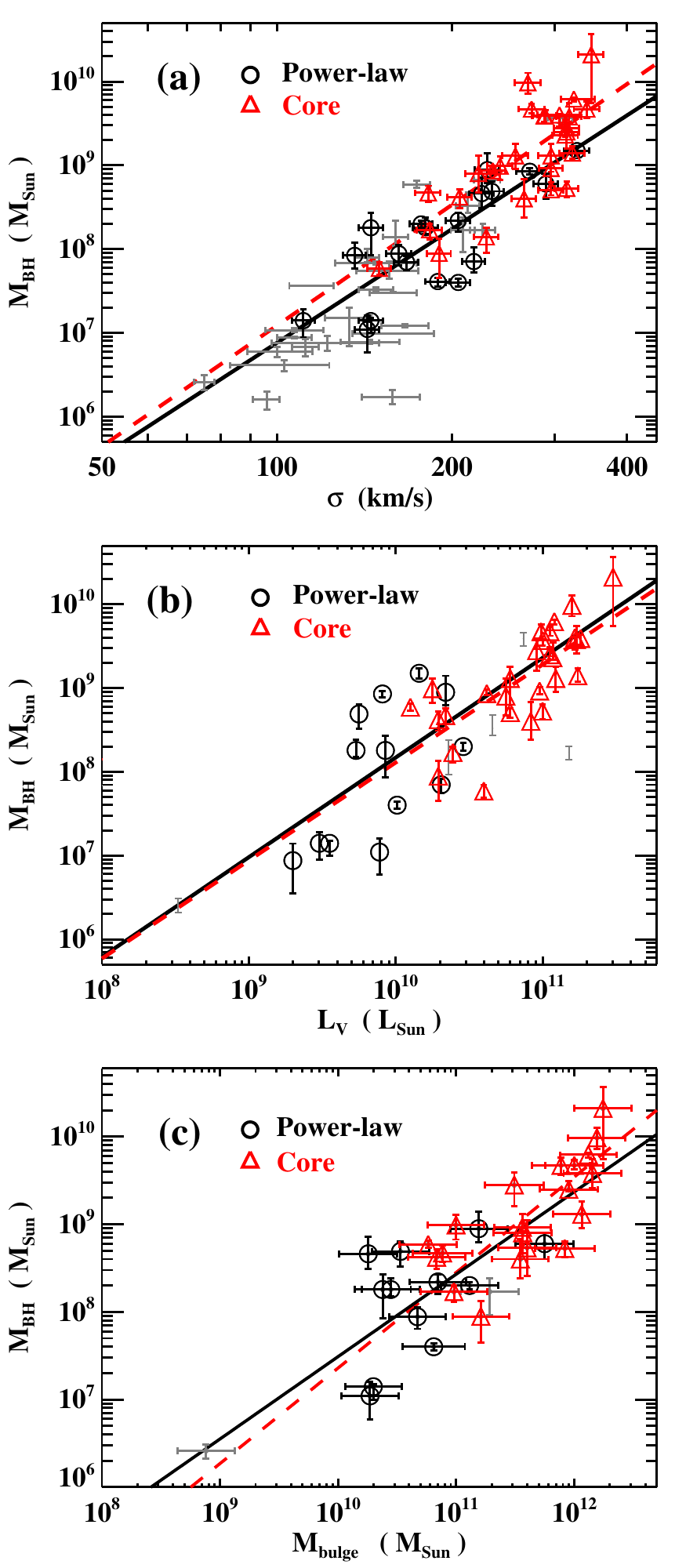,width=3.2in}
  \caption{Black hole scaling relations, with separate fits for power-law galaxies (solid black lines), versus core galaxies (dashed red lines).   
(a) $\mbh-\sigma$ relation.  (b) $\mbh-L$ relation.  (c) $\mbh-\mbulge$ relation.}
\label{fig:corepl}
\end{figure}

We also consider two subsamples of early-type galaxies classified by the
slopes of their inner surface brightness profiles, $\gamma = - d \, {\rm
  log}I / d\, {\rm log} r$.  \citet{Faber97} and \citet{Lauer07b}
distinguished ``power-law'' galaxies with $\gamma > 0.5$ from ``core''
galaxies with $\gamma < 0.3$, although other studies have reported a
continuous trend in $\gamma$ \citep[e.g.,][]{Ferr06,Glass11}.  Core
galaxies tend to be more massive and luminous than power-law galaxies, and
there is some evidence that $\mbh$ correlates with properties of the inner
stellar core \citep{Lauer07,KB09}.  In our fits to $\mbh(\sigma)$, core
galaxies have a significantly higher intercept than power-law
galaxies (see Figure~\ref{fig:corepl}a): $\alpha=8.53 \pm 0.11$ versus $8.24 \pm 0.09$.  
Our fits predict $M_{\bullet, \rm core} \sim 2 \, M_{\bullet, \rm pl}$ at $\sigma \sim 200 \kms$, where the two populations overlap.  
The offset in intercepts plus the shallower slopes ($\beta \approx$ 4.5-4.8) for core and power-law galaxies combine to produce a steeper slope ($\beta \approx 5.2$) for the early-type $\mbh-\sigma$ relation.

%

\subsubsection{Definitions of $\sigma$}

As discussed in \S\ref{sec:newsample}, the value of $\sigma$ for each
galaxy used in the $\mbh-\sigma$ relation depends on the spatial extent of
the kinematic data.  Excluding data within $\rinf$ (when resolved) has the
effect of decreasing $\sigma$ and increasing the slope of the $\mbh-\sigma$
relation.  We have obtained new values of $\sigma$ for 12 galaxies in our sample by examining kinematic profiles from the literature and excluding data within $\rinf$; these galaxies are listed in Table~\ref{tab:rinf}.  Ten of the 12 galaxies have $\sigma > 250 \kms$ using either definition.

In Table~\ref{tab:fits} we test how much the definition of $\sigma$ affects our fit to our full sample of 72 galaxies, as well as subsamples dominated by massive early-type galaxies (see rows labeled ``0-$\reff$'').
We find the slope of the global $\mbh-\sigma$ relation to change slightly from $\beta=5.64 \pm 0.32$
in our fiducial sample (in which $\sigma(\rinf-\reff)$ is used) to
$\beta=5.48 \pm 0.30$ for the conventional definition of $\sigma$ with
$r_{\rm min} = 0$ in Equation~(\ref{eq:sigeff}).  The latter is a fairer
quantity to be compared with earlier studies, but the resulting
$\mbh-\sigma$ relation is still significantly steeper than those reported
in G09 and B12.  The definition of $\sigma$ does not significantly affect our
measurements of the intrinsic scatter in log($\mbh$) (see Table~\ref{tab:fits} and \S4).

\subsubsection{High vs Low $\sigma$}

To search for possible systematic deviations of the $\mbh-\sigma$ relation
from a single power law, we divide the galaxies into low-$\sigma$ and
high-$\sigma$ subsamples, separated by a cutoff value $\scut$.  
We have tested numerous values of $\scut$ in search of robust trends.
Our strongest finding is that the relation for the higher-$\sigma$ sample appears to steepen drastically when $\scut \gtsim 270 \kms$.
As shown in Table~\ref{tab:fits} (for $\scut = 275 \kms$), the {\tt MPFITEXY} and {\tt LINMIX\_ERR} procedures both return nearly vertical relations, with very large uncertainties in $\beta$.   
This suggests a breakdown of the $\mbh-\sigma$ correlation as the galaxy population ``saturates'' at $\sigma \sim 350 \kms$.  Saturation of the $L-\sigma$ and $\mbh-\sigma$ relations has been predicted from observations and simulations of the most massive galaxies \citep[e.g.,][]{BKMaQ06,Bernardi07,Lauer07}.
The {\tt MPFITEXY} procedure returns zero intrinsic scatter when fitting galaxies with $\sigma > 275 \kms$; this is an artifact of the large slope. 

One might suspect that the saturation in $\mbh(\sigma)$ results from the lower $\sigma$ values we obtained for 12 galaxies after excluding data within $\rinf$.  We have also fit the high-$\sigma$ and low-$\sigma$ subsamples using the conventional definition of $\sigma$.  We still find that the highest-$\sigma$ galaxies follow a very steep $\mbh-\sigma$ relation, although this trend begins to appear at slightly higher values of $\scut$ (exemplified by $\scut = 290 \kms$ in Table~\ref{tab:fits}).

A weaker trend occurs for $\scut$ in the range 175-225 $\kms$.  Here, the higher-$\sigma$ sample exhibits a steeper slope ($\beta \sim$ 6-7) than the lower-$\sigma$ sample ($\beta \sim 5$).  Still, the differences
between the fits for the two subsamples are within $1\sigma$ error bars as
the uncertainties in $\alpha$ and $\beta$ are large. 
This trend vanishes when we adopt the conventional definition of $\sigma$ (data from $r = 0$-$\reff$).
As an example, Table~\ref{tab:fits} lists the fitting results for $\scut = 200 \kms$.  

It is tempting to fit $\mbh(\sigma)$ for narrow intervals in $\sigma$ (e.g., $150 \kms < \sigma < 200 \kms$), but the intrinsic scatter in the $\mbh-\sigma$ relation drives these samples toward an uncorrelated distribution.  For the current sample of $\mbh$, a long baseline in $\sigma$ ($\gtsim 100 \kms$) is therefore needed to determine the slope of the $\mbh-\sigma$ relation.

We have also tried fitting $\mbh(\sigma)$ for subsamples defined by cuts in
$L$ and $\mbulge$; two examples are listed in Table~\ref{tab:fits}.
Some of the high-$L$ subsamples exhibit a steep $\mbh-\sigma$ slope, but again with large uncertainties.

 \subsubsection{A Log-quadratic Fit to $\mbh-\sigma$}
\label{sec:quadratic}

In light of evidence that the $\mbh-\sigma$ relation steepens
toward high galaxy masses, we have also attempted to fit $\mbh(\sigma)$ as
a log-quadratic function in Equation~(\ref{eq:quadratic}).
The coefficients $\alpha$, $\beta$, $\beta_2$, and intrinsic scatter
$\epsilon_0$ for our 72-galaxy sample are determined from a brute-force
least-squares estimator similar to {\tt MPFITEXY}.  We find that the best-fit parameters
are $\alpha = 8.28 \pm 0.07$, $\beta = 5.76 \pm 0.34$, $\beta_2 = 1.68 \pm
1.82$, and $\epsilon_0 = 0.38$.  Uncertainties in $\alpha$, $\beta$, and
$\beta_2$ are determined by assessing the one-dimensional likelihood
function after marginalizing $\chi^2$ with respect to the other two
parameters.  We find $\beta_2 > 0$ with 82\% confidence, slightly below the
$1\sigma$ threshold for a one-sided confidence interval.

As suggested by the highly uncertain power-law slopes for high-$\sigma$ galaxies, our measurement of upward curvature in $\mbh(\sigma)$ is marginal.
Adopting a quadratic relation does not decrease the intrinsic scatter in
$\mbh$.  Our full sample updates the investigations of
Wyithe (2006a,b; 31 galaxies) and G09 (49 galaxies), who reported similar
confidence levels for a non-zero quadratic term.  At the extreme end of the local galaxy velocity
dispersion function ($\sigma \sim 400 \kms$), our best quadratic fit
predicts black hole masses $\sim 40\%$ higher than the best power-law fit.

%
\begin{table*}[!b]
\begin{center}
\caption{Power-law fits to black hole correlations}
\label{tab:fits}
\begin{tabular}[b]{lclccl}  
\hline
\hspace{0.1in} \bf{Sample} & \bf{$N_{\rm gal}$} & \bf{Method} & \bf{$\alpha$} & \bf{$\beta$} & \bf{$\epsilon_0$}  \\[4pt]
\hline
\bf{$\mbh- \sigma$ relation} &&&&& \\[4pt]
\hspace{0.25in} All galaxies & 72 & {\tt MPFITEXY} & $8.32 \pm 0.05$ & $5.64 \pm 0.32$ & 0.38 \\
\hspace{0.25in} All galaxies & 72 & {\tt LINMIX\_ERR} & $8.31 \pm 0.06$ & $5.67 \pm 0.33$ & $0.40 \pm 0.04$ \\
\hspace{0.25in} All + upper limits & 164 & {\tt ASURV} & $7.72 \pm 0.12$ & $5.37 \pm 0.62$ \\
\hspace{0.25in} All + upper limits & 164 & {\tt LINMIX\_ERR} & $8.15 \pm 0.05$ & $5.58 \pm 0.30$ & $0.43 \pm 0.04$ \\
\hspace{0.25in} All galaxies (0-$\reff$) & 72 & {\tt MPFITEXY} & $8.29 \pm 0.05$ & $5.48 \pm 0.30$ & 0.37 \\
\hspace{0.25in} G09 data (0-$\reff$) & 49 & {\tt MPFITEXY} & $8.19 \pm 0.06$ & $4.12 \pm 0.38$ & 0.39 \\
\\
\hspace{0.25in} Early-type & 53 & {\tt MPFITEXY} & $8.39 \pm 0.06$ & $5.20 \pm 0.36$ & 0.34 \\
\hspace{0.25in} Early-type (0-$\reff$) & 53 & {\tt MPFITEXY} & $8.36 \pm 0.05$ & $5.05 \pm 0.34$ & 0.33 \\
\hspace{0.25in} Late-type & 19 & {\tt MPFITEXY} & $8.07 \pm 0.21$ & $5.06 \pm 1.16$ & 0.46 \\
\\
\hspace{0.25in} Power-law & 18 & {\tt MPFITEXY} & $8.24 \pm 0.09$ & $4.51 \pm 0.73$ & 0.34 \\
\hspace{0.25in} Core & 28 & {\tt MPFITEXY} & $8.53 \pm 0.11$ & $4.79 \pm 0.74$ & 0.35 \\
\hspace{0.25in} Core (0-$\reff$) & 28 & {\tt MPFITEXY} & $8.50 \pm 0.11$ & $4.63 \pm 0.68$ & 0.34 \\
\\
\hspace{0.25in} $\sigma \leq 200 \kms$ & 35 & {\tt MPFITEXY} & $8.35 \pm 0.15$ & $5.66 \pm 0.85$ & 0.43 \\
\hspace{0.25in} $\sigma > 200 \kms$ & 37 & {\tt MPFITEXY} & $8.16 \pm 0.13$ & $6.76 \pm 0.91$ & 0.34 \\
\hspace{0.25in} $\sigma > 200 \kms$ (0-$\reff$)& 38 & {\tt MPFITEXY} & $8.26 \pm 0.12$ & $5.70 \pm 0.78$ & 0.35 \\
\\
\hspace{0.25in} $\sigma \leq 275 \kms$ & 55 & {\tt LINMIX\_ERR} & $8.33 \pm 0.07$ & $5.77 \pm 0.51$ & $0.43 \pm 0.05$ \\
\hspace{0.25in} $\sigma > 275 \kms$ & 17 & {\tt LINMIX\_ERR} & $7.00 \pm 2.42$ & $12.3 \pm 12.6$ & $0.34 \pm 0.11$ \\
\hspace{0.25in} $\sigma > 275 \kms$ & 17 & {\tt MPFITEXY} & $2.47 \pm 3.17$ & $35.8 \pm 16.5$ & N/A \\
\hspace{0.25in} $\sigma > 290 \kms$ (0-$\reff$)& 15 & {\tt LINMIX\_ERR} & $7.68 \pm 1.26$ & $7.93 \pm 5.79$ & $0.32 \pm 0.11$ \\
\\
\hspace{0.25in} $L \leq 10^{10.8} \lsun$ & 25 & {\tt MPFITEXY} & $8.37 \pm 0.08$ & $4.76 \pm 0.55$ & 0.33 \\
\hspace{0.25in} $L > 10^{10.8} \lsun$ & 19 & {\tt MPFITEXY} & $8.13 \pm 0.40$ & $7.19 \pm 2.25$ & 0.37\\
\hspace{0.25in} $L > 10^{10.8} \lsun$ (0-$\reff$) & 19 & {\tt MPFITEXY} & $8.29 \pm 0.33$ & $5.83 \pm 1.75$ & 0.36 \\
\\
\hspace{0.25in} $\mbulge \leq 10^{11.5} \msun$ & 21 & {\tt MPFITEXY} & $8.40 \pm 0.09$ & $5.08 \pm 0.70$ & 0.34 \\
\hspace{0.25in} $\mbulge > 10^{11.5} \msun$ & 14 & {\tt MPFITEXY} & $8.52 \pm 0.47$ & $4.69 \pm 2.69$ & 0.46\\
\hspace{0.25in} $\mbulge > 10^{11.5} \msun$ (0-$\reff$) & 14 & {\tt MPFITEXY} & $8.61 \pm 0.40$ & $3.80 \pm 2.09$ & 0.45\\[4pt]
\hline   
{\bf $\mbh- L$ relation} &&&&& \\[4pt]
\hspace{0.25in} Early-type galaxies & 44 & {\tt MPFITEXY} & $9.23 \pm 0.10$ & $1.11 \pm 0.13$ & 0.49 \\
\hspace{0.25in} Early-type galaxies & 44 & {\tt LINMIX\_ERR} & $9.23 \pm 0.10$ & $1.11 \pm 0.14$ & $0.52 \pm 0.06$ \\
\hspace{0.25in} G09 data (early-type)& 32 & {\tt MPFITEXY} & $9.01 \pm 0.10$ & $1.17 \pm 0.12$ & 0.36 \\
\\
\hspace{0.25in} Power-law & 12 & {\tt MPFITEXY} & $9.36 \pm 0.72$ & $1.19 \pm 0.67$ & 0.68 \\
\hspace{0.25in} Core & 27 & {\tt MPFITEXY} & $9.28 \pm 0.09$ & $1.17 \pm 0.22$ & 0.39 \\
\\
\hspace{0.25in} $L \leq 10^{10.8} \lsun$ & 25 & {\tt MPFITEXY} & $9.10 \pm 0.23$ & $0.98 \pm 0.20$ & 0.54 \\
\hspace{0.25in} $L > 10^{10.8} \lsun$ & 19 & {\tt MPFITEXY} & $9.27 \pm 0.13$ & $1.12 \pm 0.82$ & 0.47 \\
\\
\hspace{0.25in} $\mbulge \leq 10^{11.5} \msun$ & 18 & {\tt MPFITEXY} & $9.25 \pm 0.24$ & $1.13 \pm 0.23$ & 0.47\\
\hspace{0.25in} $\mbulge > 10^{11.5} \msun$ & 13 & {\tt MPFITEXY} & $9.24 \pm 0.10$ & $2.49 \pm 0.67$ & 0.30\\[4pt]
\hline
{\bf $\mbh - \mbulge$ relation} &&&&& \\[4pt]
\hspace{0.25in} Dynamical masses & 35 & {\tt MPFITEXY} & $8.46 \pm 0.08$ & $1.05 \pm 0.11$ & 0.34 \\
\hspace{0.25in} Dynamical masses & 35 & {\tt LINMIX\_ERR} & $8.46 \pm 0.09$ & $1.05 \pm 0.12$ & $0.36 \pm 0.08$ \\
\hspace{0.25in} Stellar masses & 18 & {\tt MPFITEXY} & $8.56 \pm 0.10$ & $1.34 \pm 0.15$ & 0.17 \\
\\
\hspace{0.25in} Power-law  & 12 & {\tt MPFITEXY} & $8.43 \pm 0.20$ & $0.94 \pm 0.39$ & 0.50 \\
\hspace{0.25in} Core    & 20 & {\tt MPFITEXY} & $8.45 \pm 0.15$ & $1.09 \pm 0.20$ & 0.28 \\
\\
\hspace{0.25in} $L \leq 10^{10.8} \lsun$ & 19 & {\tt LINMIX\_ERR} & $8.44 \pm 0.14$ & $1.05 \pm 0.26$ & $0.45 \pm 0.13$ \\
\hspace{0.25in} $L > 10^{10.8} \lsun$ & 12 & {\tt LINMIX\_ERR} & $7.66 \pm 1.60$ & $1.92 \pm 1.72$ & $0.38 \pm 0.19$ \\
\hspace{0.25in} $L > 10^{10.8} \lsun$ & 12 & {\tt MPFITEXY} & $6.92 \pm 1.05$ & $2.72 \pm 1.12$ & N/A \\
\\
\hspace{0.25in} $\mbulge \leq 10^{11.5} \msun$ & 21 & {\tt LINMIX\_ERR} & $8.54 \pm 0.15$ & $1.11 \pm 0.28$ & $0.47 \pm 0.12$ \\
\hspace{0.25in} $\mbulge > 10^{11.5} \msun$ & 14 & {\tt LINMIX\_ERR} & $7.28 \pm 1.19$ & $2.26 \pm 1.33$ & $0.30 \pm 0.17$  \\
\hspace{0.25in} $\mbulge > 10^{11.5} \msun$ & 14 & {\tt MPFITEXY} & $7.03 \pm 0.78$ & $2.53 \pm 0.85$ & N/A  \\
\hline
\end{tabular}
\end{center}
Notes: For the $\mbh-\sigma$ relation, we fit $\log_{10}(\mbh) = \alpha +
\beta \log_{10}(\sigma/200 \kms)$.  Subsamples designated (0-$\reff$) define
$\sigma$ using kinematic data over the interval $0 < r < \reff$.  For all
other subsamples, we define $\sigma$ using data over the interval $\rinf <
r < \reff$.  For the $\mbh-L$ relation, we fit $\log_{10}(\mbh) = \alpha +
\beta \log_{10}(L/10^{11} \lsun)$.  Luminosities are in $V$-band.  For the
$\mbh-\mbulge$ relation, we fit $\log_{10}(\mbh) = \alpha + \beta
\log_{10}(\mbulge/10^{11} \msun)$.  All fits except for the ``stellar
masses'' line use the sample of bulges with dynamical masses.  
\end{table*}
%

\subsection{$\mbh - L$ and $\mbh - \mbulge$ Relations}
\label{sec:mlbulge}

In Table~\ref{tab:sample}, we present $V-$band
luminosities for 44 galaxies and dynamically measured bulge masses for 35 galaxies.  
For several of the late-type galaxies in our sample, the literature contains one or more estimates of the bulge-to-total light ratio. 
Rather than judging between the various estimates, we present results for the early-types only.  
Our best fit values are $\beta=1.11 \pm 0.13$ and
$\alpha=9.23 \pm 0.10$ for the $\mbh-L$ relation, and $\beta=1.05 \pm 0.11$
and $\alpha=8.46 \pm 0.08$ for the $\mbh-\mbulge$ relation.

Additional fits to subsamples of these galaxies are listed in
Table~\ref{tab:fits}.   The $\mbh-L$ and $\mbh-\mbulge$ relations do not
show statistically significant differences between core and power-law
galaxies (see also Figures~\ref{fig:corepl}b and~\ref{fig:corepl}c).

Figures~\ref{fig:plotml} and~\ref{fig:plotmm} show that our $\mbh-L$ and
$\mbh-\mbulge$ samples both appear to have a central knot, where black holes with $10^8 \msun < \mbh < 10^9
\msun$ exhibit relatively weak correlation with $L$ or $\mbulge$.  This
feature makes it difficult to interpret the fits to high-$L$ and low-$L$
(or high-$\mbulge$ and low-$\mbulge$) subsamples.  We find tentative
evidence that the most luminous and massive galaxies ($L > 10^{10.8} \lsun$; $\mbulge > 10^{11.5} \msun$) have steeper slopes in $\mbh(L)$ and $\mbh(\mbulge)$, as exemplified in Table~\ref{tab:fits}.  Both samples are
sparsely populated at the low-$\mbh$ end.

\subsection{Comparison to Previous Studies}
\label{sec:compare}

The slope of the $\mbh-\sigma$ relation reported in prior studies has
wavered between $\sim 4$ (e.g., Gebhardt et al. 2000; Tremaine et al. 2002;
G09; B12)
and $\sim 5$ \citep[e.g.,][]{Ferr00,MF01,Graham11}.  Our best-fit slope for
the global $\mbh-\sigma$ relation falls at the steep end of this
distribution, while various subsamples exhibit a wider range of slopes ($\beta \approx 3.8$ to $\beta > 12$).  
In particular, the $\mbh-\sigma$ relation for our full sample is significantly steeper than those reported in G09 ($\beta=4.24\pm 0.41$) and B12 ($\beta=4.42\pm 0.30$).  This steepening has occurred because the newest measurements of $\mbh$ in early-type galaxies (higher $\sigma$) mostly fall above the global $\mbh-\sigma$ relation, and the newest measurements of $\mbh$ in late-type galaxies (lower $\sigma$) mostly fall below the global relation.  In addition to the significant discrepancy between the two subsamples' best-fit intercepts, both the early- and late-type $\mbh-\sigma$ relations have steepened.

Our fit to early-type galaxies is significantly steeper than the early-type
fit by G09 ($\beta = 5.20 \pm 0.36$, versus $\beta = 3.96 \pm 0.42$).  This
difference largely is due to several updates to the high-$\sigma$ galaxy sample:
new measurements of $\mbh \sim 10^{10} \msun$ in the brightest
cluster galaxies NGC 4889 and NGC 3842 \citep{mcconnell11b}, 
new measurements of $\mbh > 10^9 \msun$ in seven more galaxies \citep{rusli11,RusliThesis}, and updated
measurements increasing $\mbh$ in M87 and M60 \citep{GT09,SG10}.
Defining $\sigma$ to exclude the black hole radius of influence further steepens the early-type galaxy sample by a small amount.   
If we exclude the recent additions by \citet{mcconnell11b,mcconnell11a,mcconnell12} and \citet{RusliThesis}, we obtain $\beta = 4.77 \pm 0.36$ for 42 early-type galaxies.  Removing M87 and M60 further reduces $\beta$ to $4.55 \pm 0.37$; in addition to revised black hole masses, these two galaxies exhibit some of the largest differences in $\sigma$ in Table~\ref{tab:rinf}.

Our fit for late-type galaxies is slightly steeper than G09 ($\beta = 5.06 \pm
1.16$, versus $\beta = 4.58 \pm 1.58$).  This arises primarily from our
exclusion of NGC 1068 and NGC 2748.

Our earlier compilation of a similar sample of 67
galaxies \citep{mcconnell11b} gave $\alpha = 8.28 \pm 0.06$ and $\beta = 5.13 \pm 0.34$ for the
$\mbh-\sigma$ relation.  Our $\mbh(\sigma)$ fit to the present sample of 72
galaxies has a steeper slope of $5.64 \pm 0.32$, largely due to the
exclusion of NGC 7457, which had the lowest velocity dispersion ($\sigma =
67 \kms$) of all galaxies in the previous sample (see
\S~\ref{sec:newsample} and Gebhardt et al. 2003 for discussion of this
galaxy's central massive object).  
If we include NGC 7457 in our present sample, we obtain $\alpha = 8.33 \pm 0.05$, $\beta = 5.42 \pm 0.31$, and $\epsilon_0 = 0.40$, closer to our earlier results.

Our $\mbh-L$ and $\mbh-\mbulge$ slopes are consistent with a number of
previous investigations, including multiple bandpasses for $L$
(e.g., Marconi \& Hunt 2003; H\"{a}ring \& Rix 2004; McLure \& Dunlop 2004; G09; Schulze \& Gebhardt 2011).
For the $\mbh-L$ relation, \citet{Sani11} report different $\mbh-L_{3.6\mu \rm m}$ and $\mbh-L_V$
slopes and suggest that color corrections and extinction may be responsible
for the difference.  Their $\mbh-L_{3.6\mu \rm m}$ slope is $0.93\pm 0.10$, while
their $\mbh-L_V$ slope ranges from 1.11 to 1.40 depending on the regression
method.  These slopes are consistent with our $\mbh-L_V$ slope of $1.11 \pm 0.13$.

For the $\mbh-\mbulge$ relation, the latest compilation of 46 galaxies by B12
gives a slope of $0.79\pm 0.26$.  
We note that their $\mbulge$ values are virial estimates based on the
galaxies' $\sigma$ and $\reff$ via $\mbulge = 5.0 \sigma^2 \reff / G$.  In
comparison, our $\mbulge$ values use the mass-to-light ratios obtained from
dynamical models.

Recently, \citet{Graham12} examined the $\mbh-\sigma$ and $\mbh-\mbulge$
relations with separate fits to core and non-core galaxies, based on the
galaxy sample of \citet{HRix} and updated black hole masses from
\citet{Graham11}.  The non-core galaxies were found to follow a very steep
$\mbh-\mbulge$ relation ($\beta \sim 2$), and there was virtually no
difference in the $\mbh-\sigma$ relations for core versus non-core
galaxies.  Our relative trends for core and power-law galaxies differ from those in \citet{Graham12}.
This is likely due to differences in the galaxy samples:
our core galaxies include
eleven galaxies with $\mbh > 10^9 \msun$ that are absent from the sample used by \citet{Graham12}.
Our photometric classification of
galaxies also differs from \citet{Graham12}.  In particular, we classify
the high-$\mbulge$ object NGC 6251 as a power-law galaxy, based on the
surface brightness profile of \citet{FF99}.  Excluding NGC 6251, we measure
$\beta \approx 1.6$
for power-law galaxies on the $\mbh - \mbulge$
relation.

\section{Scatter in Black Hole Mass}
\label{sec:scatter}

For a given black hole scaling relation, the differences between the measured
values of $\mbh$ and the mean power-law relation are conventionally
interpreted as a combination of measurement errors and intrinsic scatter.
We assume the scatter in $\mbh$ to be log normal, and define the intrinsic
scatter term $\epsilon_0$ such that 
\beq 
   \chi^2 = \sum_i \, \frac{\left[
    \log_{10} \left( M_{\bullet,i}\right) - \alpha - \beta x_i
  \right]^2}{\epsilon_0^2 + \epsilon_{M,i}^2 + \beta^2 \epsilon_{x,i}^2}
\;\; ,
\label{eq:chi2}
\eeq
where $x = \log_{10}(\sigma / 200 \kms)$ for the $\mbh-\sigma$ relation, 
$x = \log_{10}(L/10^{11}\lsun)$ for the $\mbh-L$ relation, and\\
$x = \log_{10}(\mbulge/10^{11}\msun)$ for the $\mbh-\mbulge$ relation.
Here, $\epsilon_M$ is the 1-$\sigma$ error in $\log_{10}(\mbh)$, and $\epsilon_x$
is the 1-$\sigma$ error in $x$.  For a given sample and power-law fit, we
adopt the value of $\epsilon_0$ for which $\chi^2_\nu = 1$ ($\chi^2 =
N_{\rm dof}$).  G09 tested several forms of intrinsic scatter in $\mbh$ and
found log normal scatter to be an appropriate description.

Fitting the full galaxy sample for each scaling relation, we find the
intrinsic scatter in $\log_{10}(\mbh)$ to be $\epsilon_0 = 0.38$ for the
$\mbh$-$\sigma$ relation, $\epsilon_0 = 0.49$ for the $\mbh$-$L$ relation,
and $\epsilon_0 = 0.34$ for the $\mbh$-$\mbulge$ relation 
(or $\epsilon_0 = 0.17$ for $\mbh$ versus stellar $\mbulge$).  
While it is tempting to conclude that $\mbulge$ is the superior predictor of $\mbh$,
the relative errors in $\mbulge$, $\sigma$, and $L$ demand a more cautious
interpretation.  As noted in \S\ref{sec:newsample}, we have assumed that
all $\mbulge$ values have an error of at least 0.24 dex.  We have repeated
our fits to $\mbh(\mbulge)$ with a minimum error of only 0.09 dex.  Fitting
the full $\mbulge$ sample with this reduced error in $\mbulge$, we obtain a
larger intrinsic scatter  
($\epsilon_0 = 0.39$) as expected from Equation~\ref{eq:chi2}, 
while the slope and intercept of the fit do not change significantly.
Similarly, our measurements of $\epsilon_0$ for the
$\mbh-\sigma$ and $\mbh-L$ relations depend in part upon the assumed errors
in $\sigma$ ($\geq 5\%$, following G09) and $L$ (typically $< 0.05$ dex). 
In addition to evaluating $\epsilon_0$, \citet{Novak06} used earlier datasets to assess which correlation yielded the lowest predictive uncertainty in $\mbh$, given a set of host galaxy properties with measurement errors.  They found the $\mbh-\sigma$ relation to be marginally favorable for predicting black holes with $\mbh \sim 10^8 \msun$, but noted that uncertainties in the relations' slopes complicated predictions near the extrema of the relations.  Our global $\mbh-\sigma$ relation has a steeper slope ($\beta = 5.64$) than the samples evaluated by \citet{Novak06}, with $\beta$ from 3.69 to 4.59.

%
\begin{figure*}[!t]
 \centering
  \epsfig{figure=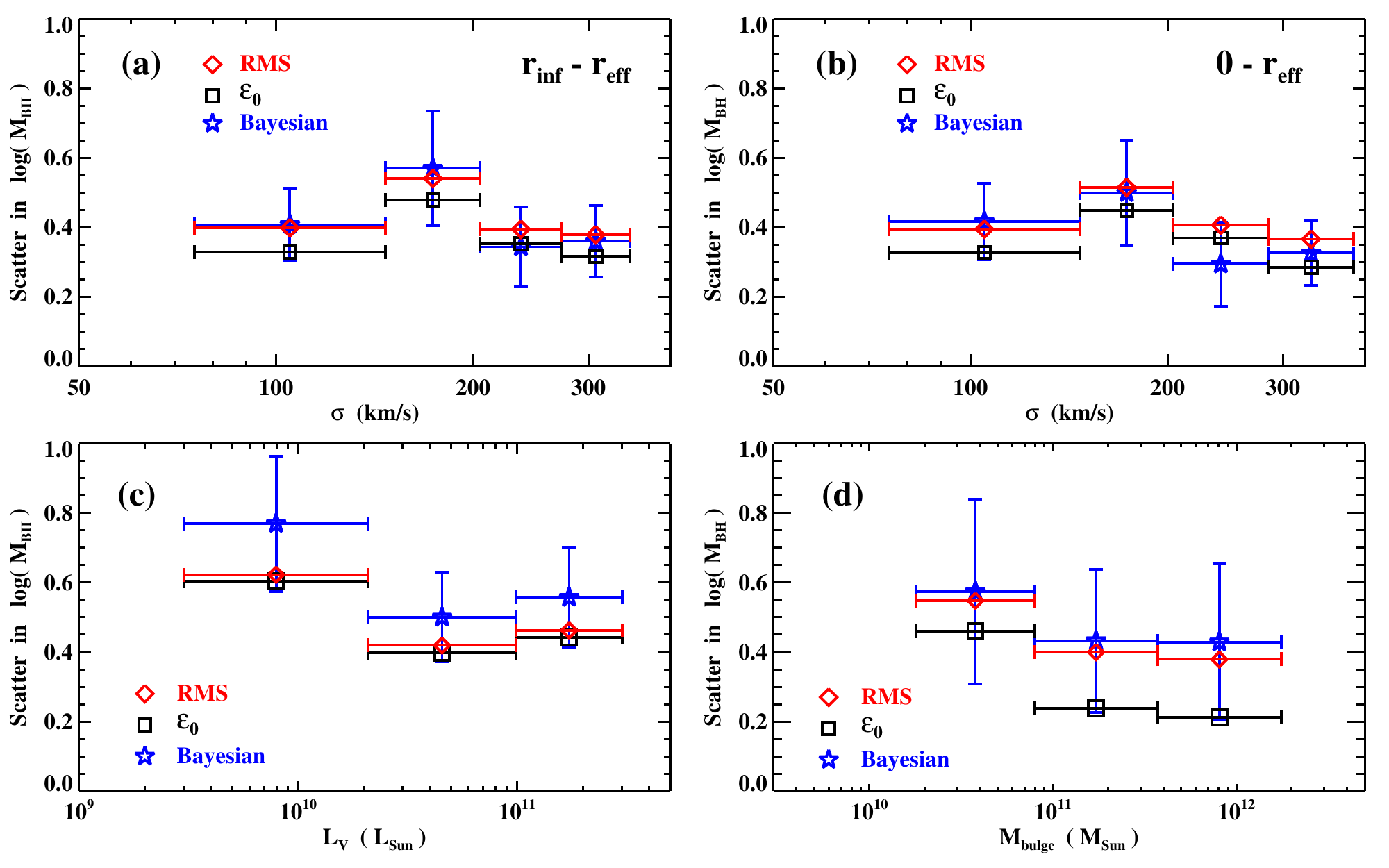,width=7.0in}
  \caption{Scatter in $\log_{10} \mbh$ for different intervals in $\sigma$, $L$,
    and $\mbulge$.  
    Black squares represent the intrinsic scatter, $\epsilon_0$ required to obtain $\chi^2 = N_{\rm gal}$ between the subsample
    of galaxies and the global scaling relation.  Red diamonds represent the root-mean-squared
    residual for each interval, between log($\mbh$) and the global scaling relation.  Blue stars with vertical error bars represent the Bayesian estimates for $\epsilon_0$, obtained while fitting a separate scaling relation for each interval.  
   (a) Scatter with respect to
    the $\mbh-\sigma$ relation, $\log_{10} \mbh = 8.32 + 5.64\, \log_{10}(\sigma/200
    \kms)$, defining $\sigma$ with data from $\rinf$ to $\reff$.  
    (b) Scatter with respect to
    the $\mbh-\sigma$ relation, $\log_{10} \mbh = 8.29 + 5.48\, \log_{10}(\sigma/200
    \kms)$, defining $\sigma$ with data from 0 to $\reff$. 
    (c) Scatter with respect to the $\mbh-L$ relation, $\log_{10} \mbh
    = 9.23 + 1.11\, \log_{10} (L/10^{11} \lsun)$.  
    (d) Scatter with respect to
    the $\mbh-\mbulge$ relation, $\log_{10} \mbh = 8.46 + 1.05\,\log_{10}(\mbulge /
    10^{11} \msun)$.  }
    \vspace{-0.1in}
\label{fig:scatter}
\vspace{0.2in}
\end{figure*}

Beyond the intrinsic scatter $\epsilon_0$ in black hole mass for the full
sample, the dependence of $\epsilon_0$ on $\sigma, L$, and $\mbulge$ is a
useful quantity for constraining theoretical models of black hole assembly.
Successive mergers are predicted to drive galaxies toward the mean
$\mbh-\mbulge$ relation, especially when black hole growth is dominated by
black hole-black hole mergers \citep[e.g.,][]{Peng07, JM11}.  Understanding
intrinsic scatter in $\mbh$ is also crucial for estimating the mass
function of black holes, starting from the luminosity function or velocity
dispersion function of galaxies (e.g., Lauer et al. 2007c; G09).

Figure~\ref{fig:scatter} illustrates how $\epsilon_0$ varies across each of
the $\mbh-\sigma$, $\mbh-L$, and $\mbh-\mbulge$ relations for our updated
sample of measurements.  For each relation, we construct three or four bins
containing equal numbers of galaxies and perform multiple estimates of the scatter in each bin.
One estimate, represented by blue stars in Figure~\ref{fig:scatter} is to perform an independent Bayesian fit (with {\tt LINMIX\_ERR}) for the scaling relation in each bin, and assess the posterior distribution of $\epsilon_0$.  This method provides uncertainties for $\epsilon_0$ in each bin, but the fits to narrow data intervals typically yield poor estimates for the slope and intercept.
A second estimate, represented by black squares in Figure~\ref{fig:scatter}, is to compute $\epsilon_0$ as in Equation~\ref{eq:chi2}, using the global fit to the scaling relation to define the same values of $\alpha$ and $\beta$ are for all bins.  
The third and simplest estimate, represented by red diamonds in Figure~\ref{fig:scatter}, is to evaluate the root-mean-squared (RMS) residual between log($\mbh$) and the global scaling relation.  
The $\epsilon_0$ term in Equation~\ref{eq:chi2} provides a reliable assessment of intrinsic scatter
only if random measurement errors are small: measurements with large
uncertainties can yield $\chi^2_\nu \leq 1$ with no intrinsic scatter term. 
In comparison, the RMS estimate has no explicit dependence on measurement errors.

Figures~\ref{fig:scatter}a and~\ref{fig:scatter}b illustrate the
  scatter in $\mbh$ as a function of $\sigma$.  Two definitions of $\sigma$
  for the 12 galaxies in Table~\ref{tab:rinf} are shown for comparison,
  where $\sigma$ is computed from data between radii $\rinf$ and $\reff$,
  or between 0 and $\reff$.  Considering the large error bars in
  $\epsilon_0$ from the Bayesian fits, we find no significant variation in
  $\mbh$ with respect to $\sigma$.  The high- and low-mass ends of the
  $\mbh-\sigma$ relation both exhibit $\sim 0.3$-0.4 dex of scatter,
  regardless of how $\sigma$ is defined or how scatter is estimated.
 
For the $\mbh-L$ and $\mbh-\mbulge$ relations, we find possible
evidence that galaxies with low spheroid luminosities ($L < 10^{10.3}
\lsun$) and small stellar masses ($\mbulge < 10^{11} \msun$) exhibit increased
intrinsic scatter in $\mbh$.  Yet the scatter appears constant for galaxies
above this range, spanning $10^{10.3} \lsun < L < 10^{11.5} \lsun$ and
$10^{11} \msun < \mbulge < 10^{12.3} \msun$.  
More measurements in the range $\mbulge \sim 10^8 - 10^{10} \msun$ are
needed to reveal whether intrinsic scatter in $\mbh$ varies systematically
across an extended range of bulge luminosities or masses.  The Bayesian estimates of $\epsilon_0$ in each bin have large uncertainties; adopting this method, we do not detect a significant change in scatter for any interval in $\sigma$, $L$, or $\mbulge$.

The Local Group galaxy M32 is separated from the other early-type galaxies in our sample by almost an order of magnitude in $L$, and more than an order of magnitude in $\mbulge$.  We have excluded its contribution in 
Figures~\ref{fig:scatter}c and~\ref{fig:scatter}d, 
so that the sizes of the leftmost bins better reflect the sampled distributions of $L$ and $\mbulge$.  Including M32 does not substantially change the amount of scatter in the lowest-$L$ and lowest-$\mbulge$ bins.

Although our three estimates of scatter do not yield the same absolute values, their qualitative trends as a function of $\sigma$, $L$, or $\mbulge$ are very similar.  The similar behavior of $\epsilon_0$ and RMS indicates that variations in measurement errors are not responsible for the apparent trends in intrinsic scatter.

\section{Summary and Discussion}
\label{sec:disc}

We have compiled an updated sample of 72 black hole masses and host galaxy
properties in Table~\ref{tab:sample}; a more detailed version of
Table~\ref{tab:sample} is available at {\tt
  http://blackhole.berkeley.edu}.  Compared with the 49 objects in G09, 27
black holes in our sample of 72 are new measurements and 18 masses are
updates of previous values from improved data and/or modeling.  
Our present sample includes updated distances to 44 galaxies.

We have presented revised fits for the $\mbh-\sigma$, $\mbh-L$, and
$\mbh-\mbulge$ relations of our updated sample (Table~\ref{tab:fits} and
Figures~\ref{fig:plotmsig}-\ref{fig:plotmm}).  Each relation is fit as a
power-law: $\log_{10}(\mbh) = \alpha + \beta \, \log_{10}(X)$.  Our best
fit to the full sample of 72 galaxies with velocity dispersion measurements
($X \equiv \sigma/200 \kms$) is $\alpha = 8.32 \pm 0.05$ and $\beta = 5.64
\pm 0.31$.  
A quadratic fit to the $\mbh-\sigma$ relation with an additional
term $\beta_2 \, [\log_{10}(X)]^2$ gives $\beta_2=1.68 \pm 1.82$
and does not decrease the intrinsic scatter in $\mbh$.
Including 92 additional upper limits for $\mbh$ decreases the intercept but does not change
the slope:
$\alpha = 8.15 \pm 0.05$ and 
$\beta = 5.58 \pm 0.30$.  

For the 44 early-type galaxies with reliable $V$-band luminosity
measurements ($X \equiv L_V/10^{11}\lsun$), we find $\alpha = 9.23 \pm
0.10$ and $\beta = 1.11 \pm 0.13$.  For the 35 early-type galaxies with
dynamical measurements of the bulge stellar mass ($X \equiv
\mbulge/10^{11}\msun$), we find $\alpha = 8.46 \pm 0.08$ and $\beta = 1.05
\pm 0.11$.  

We have also examined the black hole scaling relations for different
subsamples of galaxies.  When the galaxies are separated into early- and
late-types and fit individually for the $\mbh-\sigma$ relation, we find
similar slopes of $\beta = 5.20 \pm 0.36$ (early-types) and $5.06
\pm 1.16$ (late-types).  The intercepts, however, differ significantly:
$\alpha = 8.39 \pm 0.06$ for the early types, a factor of $\sim 2$ higher
than $\alpha=8.07 \pm 0.21$ for the late types.  The steep global slope of
5.64 is therefore largely an effect of combining different galaxy types,
each of which obey a shallower $\mbh-\sigma$ relation and different
intercepts.

When the early-type galaxies are further divided into two subsamples based
on their inner surface brightness profiles, the resulting $\mbh-\sigma$
relation has a significantly larger intercept for the core
galaxies than the power-law galaxies (Table~\ref{tab:fits} and Figure~\ref{fig:corepl}a).  The slopes of the
$\mbh-L$ and $\mbh-\mbulge$ relations do not show statistically significant
differences between core and power-law galaxies, but $\mbh$ follows $L$ and
$\mbulge$ more tightly in core galaxies than power-law galaxies (Table~\ref{tab:fits}).

In the literature, the exact value of the $\mbh-\sigma$ slope has been much
debated by observers and regarded by theorists as a key discriminator for
models of the assembly and growth of supermassive black holes and their
host galaxies. We suggest that the individual observed $\mbh-\sigma$
relations for the early- and late- type galaxies provide more meaningful
constraints on theoretical models than the global relation.  After all,
these two types of galaxies are formed via different processes.
Superficially, our measurement of $\beta = 5.64 \pm 0.32$ for the global
$\; \mbh-\sigma$ relation favors thermally-driven wind models that predict
$\beta \sim 5$ \citep[e.g.,][]{Silk98} over momentum-driven wind models
with $\beta \sim 4$ \citep[e.g.,][]{Fabian99}.  
However, the intercept of the empirical $\mbh-\sigma$ relation is substantially higher than intercepts derived from thermally-driven wind models
\citep[e.g.,][]{King10a,King10b}.

For the subsamples, the early-type
galaxies give $\beta > 4.0$ with 99.96\% confidence ($\Delta\chi^2 = 11.3$
for $\epsilon_0 = 0.34$) and $\beta > 4.5$ with 97\% confidence
($\Delta\chi^2 = 3.8$), whereas the late-type and power-law galaxy
subsamples are each consistent with $\beta = 4.0$ ($\Delta\chi^2 < 1$).  
Core galaxies exceed $\beta = 4.0$ with marginal significance ($\Delta\chi^2 = 1.1$) and are consistent with $\beta = 4.5$.
Including central kinematics in the definition of $\sigma$ further erodes
the significance of high $\beta$ for the early-type and core galaxy
subsamples.  More robust black hole measurements and more sophisticated
theoretical models taking into account of galaxy types and environment \citep[e.g.,][]{ZubovasKing12} 
are needed before stronger constraints can be obtained.

The intrinsic scatter in $\mbh$ plotted in Figure~\ref{fig:scatter} for
different intervals of $\sigma$, $L$, and $\mbulge$ serves as an
independent test for theoretical models of black hole and galaxy
growth.  Our dataset shows decreasing scatter in $\mbh$ with increasing $\sigma$.  
However, there are currently insufficient data to probe the $30 - 100 \kms$ range, where scatter in
$\mbh$ could identify the initial formation mechanism for massive black
holes \citep{VLN08,VN09}.  Theoretical models of hierarchical mergers in
$\Lambda$CDM cosmology predict that scatter in $\mbh$ should decline
steadily with increasing stellar mass ($\mstar$), even when $\mbh$ and
$\mstar$ are initially uncorrelated \citep{Malbon,Peng07,Hirsch10,JM11}.
The semi-analytic models by \citet{Malbon}, for instance, predict that
black holes with present-day masses $> 10^8 \msun$ have gained most of
their mass via black hole-black hole mergers, yielding extremely low
scatter ($\epsilon_0 \sim 0.1$) at the upper end of the $\mbh - \mbulge$
relation.  More recent models by \citet{JM11} use fully decoupled
prescriptions for star formation and black hole growth, and attain a more
realistic amount of scatter on average.  Yet these models still exhibit
decreasing scatter as $\mbulge$ increases from $\sim 10^9 \msun$ to $\sim
10^{11.5} \msun$.  In comparison, we observe nearly constant scatter from
$\mbulge \sim 10^{11} \msun$ to $10^{12} \msun$, beyond the highest bulge
masses produced in the \citet{JM11} models.

Our final comment is that investigations using the $\mbh-\sigma$
correlation should consider the definition of $\sigma$, i.e., whether it is
measured from an inner radius of zero or $\rinf$.  We find that both
definitions yield similar amounts of scatter in the $\mbh-\sigma$ relation
(Table~\ref{tab:fits}), so neither has a clear advantage for predicting
$\mbh$.  Excluding data within $\rinf$ corresponds more closely to cases
where $\rinf$ is unresolved, such as seeing-limited galaxy surveys,
high-redshift observations, or numerical simulations with limited spatial
resolution.  From a theoretical perspective, the evolutionary origin of an
$\mbh-\sigma$ relation and the immediate effects of gravity may warrant
separate consideration.  On the other hand, the total gravitational
potential of a galaxy includes its black hole.  Our test of how redefining
$\sigma$ alters the $\mbh-\sigma$ relation has only considered 12
galaxies for which data within $\rinf$ contribute prominently to the
spatially integrated velocity dispersion.  At present, the full sample of
$\mbh$ and $\sigma$ measurements comprises a heterogeneous selection of
kinematic data.  Rather than advocating a particular definition, we wish to
call attention to the nuances of interpreting the $\mbh-\sigma$ relation
and encourage future investigators to consider their options carefully.

As this manuscript was being finalized, \citet{vdB12} reported a
  measurement of $\mbh = 1.7 \pm 0.3 \times 10^{10} \msun$ in NGC 1277.
  They reported $\sigma = 333 \kms$ for data between $\rinf$ and $\reff$,
  and $\mbulge = 1.2 \pm 0.4 \times 10^{11} \msun$; the latter measurement
  suggests that NGC 1277 lies two orders of magnitude above the mean
  $\mbh-\mbulge$ relation.  Adding NGC 1277 to our 72-galaxy sample changes 
  our global power-law fit to the $\mbh-\sigma$ relation only slightly:  
  $\alpha = 8.33 \pm
  0.05$, $\beta= 5.73 \pm 0.32$, and $\epsilon_0 =
  0.39$ 
(from {\tt MPFITEXY}).  
Our global fit to $\mbh(\mbulge)$ for 36 galaxies including NGC 1277 yields 
$\alpha = 8.51 \pm 0.09$, $\beta= 1.05 \pm 0.13$, and $\epsilon_0 = 0.44$.

Dynamical measurements of $\mbh$ require substantial observational
resources and careful analysis, and are often published individually.
Nonetheless, recent and ongoing efforts are rapidly expanding the available
$\mbh$ measurements and revising the empirical black hole scaling
relations.  Our online database \footnote{\tt
  http://blackhole.berkeley.edu} is aimed to provide all researchers easy
access to frequently updated compilation of supermassive black holes with
direct dynamical mass measurements and their host galaxy properties.
Updated scaling relations can be used to estimate $\mbh$ more accurately in
individual galaxies.  This can improve our knowledge of Eddington rates and
spectral energy distributions for accreting black holes, as well as time
and distance scales for tidal disruption events.  Moreover, the
$\mbh-\sigma$ relation for quiescent black holes has been used to normalize
the black hole masses obtained from reverberation mapping studies of active
galaxies \citep{Onken04,Woo10,Park12}.  This important calibration could be
improved by addressing morphology biases in the reverberation mapping
samples and the $\mbh-\sigma$ relations for different galaxy types.

\medskip

This work is supported in part by NSF AST-1009663.  NJM is supported by the Beatrice Watson Parrent Fellowship.
We thank Karl Gebhardt, Tod Lauer, and John Blakeslee for useful discussions
and Michael Reed for help with the data table and compilation.  We thank the anonymous referee for constructive comments on our original manuscript.

%
\begin{table*}
\renewcommand\thetable{A1}
\begin{center}
\caption{Galaxies with dynamical measurements of $\mbh$}
\label{tab:sample}
\begin{tabular}[b]{lcccclcclcl}  

\hline
\bf{Galaxy} & \bf{$\mbh \; (+,-)$} & \bf{Ref.} & \bf{$\sigma$} & $\log{L_V}$ & \bf{$\mbulge$} & \bf{Ref.} & \bf{$\rinf$} & \bf{Morph.} & \bf{$D$} & \bf{Method}\\
 & ($\msun$) & & ($\kms$) & & ($\msun$) & & (arcsec) & & (Mpc) & \\
\hline 
\\

Milky Way $^a$ & 4.1 (0.6,0.6) e6 & 1,2 & $103 \pm 20$ & & & & 43 & S & 0.008 & stars \\

A1836-BCG & 3.9 (0.4,0.6) e9 & 3 & $288 \pm 14$ & $11.26 \pm 0.06$ & & & 0.27 & E (C) & 157.5 & gas \\

A3565-BCG & 1.4 (0.3,0.2) e9 & 3 & $322 \pm 16$ & $11.24 \pm 0.06$ & & & 0.22 & E (C) & 54.4 & gas \\

Circinus & 1.7 (0.4,0.3) e6 & 4 & $158 \pm 18$ & & & & 0.02 & S & 4.0 & masers \\

IC 1459 $^b$ & 2.8 (1.1,1.2) e9 & 5 & $315 \pm 16$ & $10.96 \pm 0.06$ & 3.07e11 & 45 & 0.81 & E (C) & 30.9 & stars \\

N221 (M32) $^y$ & 2.6 (0.5,0.5) e6 & 6 & $75 \pm 3$ & $8.52 \pm 0.02$ & 7.62e8 & 45 & 0.57 & E (I) & 0.73 & stars \\

N224 (M31) $^y$ & 1.4 (0.8,0.3) e8 & 7 & $160 \pm 8$ & & & & 6.5 & S & 0.73 & stars \\

N524 $^w$ & 8.6 (1.0,0.4) e8 & 8 & $235 \pm 12$ & $10.62 \pm 0.04$ & & & 0.57 & S0 (C) & 24.2 & stars \\

N821 $^w$ & 1.7 (0.7,0.7) e8 & 9 & $209 \pm 10$ & $10.36 \pm 0.05$ & 1.92e11 & 9 & 0.14 & E (I) & 23.4 & stars \\ 

N1023 $^w$ & 4.0 (0.4,0.4) e7 & 10 & $205 \pm 10$ & $10.06 \pm 0.11$ & 6.49e10 & 45 & 0.08 & S0 (pl) & 10.5 & stars \\

N1194 $^c$ & 6.8 (0.3,0.3) e7 & 11 & $148^{+26}_{-22}$ & & & & 0.05 & S0 & 55.5 & masers \\

N1300 & 7.1 (3.4,1.8) e7 & 12 & $218 \pm 10$ & & & & 0.07 & S & 20.1 & gas \\

N1316 $^x$ & 1.7 (0.3,0.3) e8 & 13 & $226 \pm 11$ & $11.18 \pm 0.05$ & & & 0.14 & E (I) & 21.0 & stars \\

N1332 $^w$ & 1.5 (0.2,0.2) e9 & 14 & $328 \pm 16$ & $10.16 \pm 0.05$ & & & 0.54 & S0 (pl) & 22.7 & stars \\

N1374 $^{b,x}$ & 5.9 (0.6,0.5) e8 & 15 & $174 \pm 9$ & $10.10 \pm 0.05$ & 5.79e10 & 15 & 0.89 & E (C) & 19.6 & stars \\

N1399 $^{b,d,x}$ & 5.1 (0.6,0.7) e8 & 16 & $296 \pm 15$ & $10.78 \pm 0.04$ & 3.98e11 & 46 & 0.25 & E (C) & 20.9 & stars \\

N1399 $^{b,d,x}$ & 1.3 (0.5,0.7) e9 & 17 & $296 \pm 15$ & $10.78 \pm 0.04$ & 3.98e11 & 46 & 0.63 & E (C) & 20.9 & stars \\

N1407 $^{b,w}$ & 4.7 (0.7,0.5) e9 & 15 & $274 \pm 14$ & $11.05 \pm 0.05$ & 1.00e12 & 15 & 1.9 & E (C) & 29.0 & stars \\

N1550 $^b$ & 3.9 (0.7,0.7) e9 & 15 & $289 \pm 14$ & $10.87 \pm 0.05$ & & & 0.78 & E (I) & 53.0 & stars \\

N2273 $^c$ & 7.8 (0.4,0.4) e6 & 11 & $144^{+18}_{-15}$ & & & & 0.01 & S & 26.8 & masers \\

N2549 $^w$ & 1.4 (0.1,0.4) e7 & 8 & $145 \pm 7$ & $9.55 \pm 0.04$ & 1.99e10 & 8 & 0.05 & S0 (pl) & 12.7 & stars \\

N2787 $^w$ & 4.1 (0.4,0.5) e7 & 18 & $189 \pm 9$ & & & & 0.14 & S0 (pl) & 7.5 & gas \\

N2960 $^c$ & 1.21 (0.05,0.05) e7 & 11 & $166^{+16}_{-15}$ & & & & 0.01 & S & 75.3 & masers \\

N3031 (M81) & 8.0 (2.0,1.1) e7 & 19 & $143 \pm 7$ & & & & 0.85 & S & 4.1 & gas \\

N3091 & 3.7 (0.1,0.5) e9 & 15 & $307 \pm 15$ & $11.00 \pm 0.05$ & & & 0.66 & E (C) & 52.7 & stars \\

N3115 $^w$ & 8.9 (5.1,2.7) e8 & 20 & $230 \pm 11$ & $10.34 \pm 0.02$ & 1.57e11 & 45 & 1.6 & S0 (pl) & 9.5 & stars \\

N3227 & 1.5 (0.5,0.8) e7 & 21 & $133 \pm 12$ & & & & 0.04 & S & 17.0 & stars \\

N3245 $^w$ & 2.1 (0.5,0.6) e8 & 22 & $205 \pm 10$ & & 7.00e10 & 45 & 0.21 & S0 (pl) & 21.5 & gas \\

N3368 $^w$ & 7.6 (1.6,1.5) e6 & 23 & $122^{+28}_{-24}$ & & & & 0.04 & S & 10.6 & stars \\

N3377 $^w$ & 1.8 (0.9,0.9) e8 & 9 & $145 \pm 7$ & $9.93 \pm 0.04$ & 2.35e10 & 9 & 0.69 & E (pl) & 11.0 & stars \\

N3379 (M105) $^w$ & 4.2 (1.0,1.1) e8 & 24 & $206 \pm 10$ & $10.29 \pm 0.01$ & 6.86e10 & 45 & 0.83 & E (C) & 10.7 & stars \\

N3384 $^w$ & 1.1 (0.5,0.5) e7 & 9 & $143 \pm 7$ & $9.89 \pm 0.09$ & 1.90e10 & 9 & 0.04 & S0 (pl) & 11.5 & stars \\

N3393 & 3.3 (0.2,0.2) e7 & 25 & $148 \pm 10$ & & & & 0.03 & S & 53.6 & masers \\

N3489 $^w$ & 6.0 (0.8,0.9) e6 & 23 & $100^{+15}_{-11}$ & & & & 0.04 & S0 & 12.0 & stars \\

N3585 $^w$ & 3.3 (1.5,0.6) e8 & 26 & $213 \pm 10$ & $10.66 \pm 0.08$ & 1.60e11 & 26 & 0.31 & S0 (I) & 20.6 & stars\\

N3607 $^{e,w}$ & 1.4 (0.4,0.5) e8 & 26 & $229 \pm 11$ & & & & 0.10 & E (C) & 22.6 & stars \\

N3608 $^w$ & 4.7 (1.0,1.0) e8 & 9 & $182 \pm 9$ & $10.34 \pm 0.04$ & 7.66e10 & 9 & 0.55 & E (C) & 22.8 & stars \\

N3842 $^b$ & 9.7 (3.0,2.5) e9 & 27 & $270 \pm 14$ & $11.20 \pm 0.05$ & 1.55e12 & 44 & 1.2 & E (C) & 98.4 & stars \\

N3998 $^{b,w}$ & 8.5 (0.7,0.7) e8 & 28 & $272 \pm 14$ & $9.91 \pm 0.04$ & & & 0.71 & S0 (pl) & 14.3 & stars \\

N4026 $^w$ & 1.8 (0.6,0.3) e8 & 26 & $180 \pm 9$ & $9.73 \pm 0.08$ & 2.81e10 & 26 & 0.37 & S0 (pl) & 13.4 & stars \\

N4258 $^w$ & 3.67 (0.01,0.01) e7 & 29 & $115 \pm 10$ & & & & 0.35 & S & 7.0 & masers \\

N4261 $^w$ & 5.3 (1.1,1.1) e8 & 30 & $315 \pm 15$ & $11.00 \pm 0.02$ & 8.26e11 & 45 & 0.15 & E (C) & 32.6 & gas \\

N4291 $^w$ & 9.8 (3.1,3.1) e8 & 9 & $242 \pm 12$ & $10.25 \pm 0.05$ & 9.96e10 & 9 & 0.56 & E (C) & 26.6 & stars \\

N4342 $^z$ & 4.6 (2.6,1.5) e8 & 31 & $225 \pm 11$ & & 1.80e10 & 45 & 0.35 & S0 (pl) & 23.0 & stars \\

N4374 (M84) $^x$ & 9.2 (1.0,0.8) e8 & 32 & $296 \pm 14$ & $10.98 \pm 0.02$ & 3.62e11 & 45 & 0.51 & E (C) & 18.5 & gas \\ 

N4388 $^c$ & 8.8 (0.2,0.2) e6 & 11 & $107^{+8}_{-7}$ & & & & 0.03 & S & 19.8 & masers \\

N4459 $^x$ & 7.0 (1.3,1.4) e7 & 18 & $167 \pm 8$ & $10.31 \pm 0.02$ & & & 0.14 & E (pl) & 16.0 & gas \\

N4472 (M49) $^x$ & 2.5 (0.6,0.1) e9 & 15 & $300 \pm 15$ & $11.05 \pm 0.05$ & 8.98e11 & 15 & 1.3 & E (C) & 16.7 & stars \\

N4473 $^x$ & 8.9 (4.5,4.4) e7 & 9 & $190 \pm 9$ & $10.29 \pm 0.02$ & 1.61e11 & 9 & 0.15 & E (C) & 15.2 & stars \\

N4486 (M87) $^{b,x}$ & 6.2 (0.3,0.4) e9 & 33 & $324^{+28}_{-16}$ & $11.08 \pm 0.02$ & 1.31e12 & 47 & 3.1 & E (C) & 16.7 & stars \\

N4486A $^x$ & 1.4 (0.5,0.5) e7 & 34 & $111 \pm 5$ & $9.48 \pm 0.02$ & & & 0.06 & E (pl) & 18.4 & stars \\

N4564 $^{f,x}$ & 8.8 (2.4,2.4) e7 & 9 & $162 \pm 8$ & & 4.66e10 & 45 & 0.19 & S0 (pl) & 15.9 & stars \\

N4594 (M104) $^{b,w}$ & 6.7 (0.5,0.4) e8 & 35 & $230 \pm 12$ & & & & 1.1 & S & 10.0 & stars \\

N4596 & 8.4 (3.6,2.5) e7 & 18 & $136 \pm 6$ & & & & 0.22 & S0 (pl) & 18.0 & gas \\

N4649 (M60) $^{b,x}$ & 4.7 (1.1,1.0) e9 & 36 & $341 \pm 17$ & $10.99 \pm 0.02$ & 7.72e11 & 36 & 2.2 & E (C) & 16.5 & stars \\

N4697 $^x$ & 2.0 (0.2,0.2) e8 & 9 & $177 \pm 8$ & $10.46 \pm 0.04$ & 1.29e11 & 9 & 0.46 & E (pl) & 12.5 & stars \\

N4736 (M94) $^{g,w}$ & 6.8 (1.6,1.6) e6 & 37 & $112 \pm 6$ & & & & 0.10 & S & 5.0 & stars \\

N4826 (M64) $^{g,w}$ & 1.6 (0.4,0.4) e6 & 37 & $96 \pm 5$ & & & & 0.02 & S & 7.3 & stars \\

N4889 $^b$ & 2.1 (1.6,1.55) e10 & 27 & $347 \pm 17$ & $11.48 \pm 0.05$ & 1.75e12 & 46 & 1.5 & E (C) & 103.2 & stars \\
\hline

\end{tabular}
\end{center}
\end{table*}

\begin{table*}[!h]
\begin{center}
\textbf{Table A1, continued}\\
\medskip

\begin{tabular}[b]{lcccccccccc}  

\hline
\bf{Galaxy} & \bf{$\mbh \; (+,-)$} & \bf{Ref.} & \bf{$\sigma$} & $\log{L_V}$ & \bf{$\mbulge$} & \bf{Ref.} & \bf{$\rinf$} & \bf{Morph.} & \bf{$D$} & \bf{Method}\\
 & ($\msun$) & & ($\kms$) & & ($\msun$) & & (arcsec) & & (Mpc) & \\
\hline 
\\

N5077 & 8.0 (5.0,3.3) e8 & 38 & $222 \pm 11$ & $10.75 \pm 0.05$ & 3.66e11 & 38 & 0.32 & E (C) & 44.9 & gas \\

N5128 (Cen A) $^{h,w}$ & 5.9 (1.1,1.0) e7 & 39 & $150 \pm 7$ & $10.60 \pm 0.03$ & & & 0.60 & S0/E (C) & 4.1 & stars \\

N5516 & 4.0 (0.1,1.1) e9 & 15 & $306 \pm 26$ & $11.22 \pm 0.05$ & & & 0.63 & E (C) & 60.1 & stars \\

N5576 $^w$ & 1.7 (0.3,0.4) e8 & 26 & $183 \pm 9$ & $10.39 \pm 0.05$ & 9.58e10 & 26 & 0.18 & E (C) & 25.7 & stars \\

N5845 $^w$ & 4.9 (1.5,1.6) e8 & 9 & $234 \pm 11$ & $9.75 \pm 0.05$ & 3.36e10 & 9 & 0.31 & E (pl) & 25.9 & stars \\

N6086 & 3.8 (1.7,1.2) e9 & 40 & $318 \pm 16$ & $11.23 \pm 0.05$ & 1.43e12 & 40 & 0.24 & E (C) & 139.1 & stars \\

N6251 & 6.0 (2.0,2.0) e8 & 41 & $290 \pm 14$ & & 5.60e11 & 46 & 0.06 & E (pl) & 106.0 & gas \\

N6264 $^c$ & 3.03 (0.05,0.04) e7 & 11 & $158^{+16}_{-14}$ & & & & 0.01 & S & 145.4 & masers \\

N6323 $^c$ & 9.8 (0.1,0.1) e6 & 11 & $158^{+28}_{-23}$ & & & & 0.003 & S & 110.5 & masers \\

N7052 & 4.0 (2.8,1.6) e8 & 42 & $266 \pm 13$ & $10.92 \pm 0.04$ & 3.50e11 & 45 & 0.07 & E (C) & 70.9 & gas \\

N7582 & 5.5 (1.6,1.1) e7 & 43 & $156 \pm 19$ & & & & 0.09 & S & 22.3 & gas \\

N7619 $^{b,w}$ & 2.3 (1.2,0.1) e9 & 15 & $313 \pm 16$ & $11.07 \pm 0.05$ & & & 0.39 & E (C) & 53.9 & stars \\

N7768 $^b$ & 1.3 (0.5,0.4) e9 & 44 & $257 \pm 13$ & $11.09 \pm 0.05$ & 1.16e12 & 44 & 0.14 & E (C) & 112.8 & stars \\

U3789 $^c$ & 1.08 (0.06,0.05) e7 & 11 & $107^{+13}_{-12}$ & & & & 0.02 & S & 48.4 & masers \\

\hline

\end{tabular}
\end{center}

Notes: 
The first reference column corresponds to the black hole mass measurement, and the second corresponds to the measurement of $\ml$ used to compute $\mbulge$.  Bulge luminosity $L_V$ is in solar units.  
Quoted errors $(+,-)$ for $\mbh$ are $68\%$ confidence intervals.
We assume 0.24 dex uncertainty for all $\mbulge$ values.
The black hole radius of influence $\rinf$ is defined by $G\mbh/\sigma^2$.
Morphologies include designations for power-law (pl), core (C), and intermediate (I) surface brightness profiles.
Distances for 44 objects have been updated since the compilation of \citet{mcconnell11b}; see notes $w$-$z$ below.
A more detailed version of this table is available at {\tt http://blackhole.berkeley.edu}.\\
  
References:  \citep[1 =][]{ghez08} ; \citep[2 =][]{gillessen09} ; \citep[3 =][]{Bonta} ; \citep[4 =][]{greenhill03} ; \citep[5 =][]{Capp02} ; \citep[6 =][]{Verolme} ; \citep[7 =][]{Bender05} ; \citep[8 =][]{krajnovic09} ; \citep[9 =][]{Schulze11} ; \citep[10 =][]{bower01} ; \citep[11 =][]{Kuo11} ; \citep[12 =][]{atkinson05} ; \citep[13 =][]{Nowak08} ; \citep[14 =][]{rusli11} ; \citep[15 =][]{RusliThesis} ; \citep[16 =][]{Geb07} ; \citep[17 =][]{houghton} ; \citep[18 =][]{sarzi01} ; \citep[19 =][]{devereux03} ; \citep[20 =][]{emsellem99} ; \citep[21 =][]{davies06} ; \citep[22 =][]{barth01} ; \citep[23 =][]{Nowak10} ; \citep[24 =][]{vdB10} ; \citep[25 =][]{Kondratko} ; \citep[26 =][]{Gultekinb} ; \citep[27 =][]{mcconnell11b} ; \citep[28 =][]{walsh12} ; \citep[29 =][]{Herrnstein05} ; \citep[30 =][]{FFJ96} ; \citep[31 =][]{cretton99} ; \citep[32 =][]{walsh10} ; \citep[33 =][]{Geb11} ; \citep[34 =][]{Nowak07} ; \citep[35 =][]{Jardel11} ; \citep[36 =][]{SG10} ; \citep[37 =][]{KBC11} ; \citep[38 =][]{DeFrancesco} ; \citep[39 =][]{Capp09} ; \citep[40 =][]{mcconnell11a} ; \citep[41 =][]{FF99} ; \citep[42 =][]{vdMvdB98} ; \citep[43 =][]{wold06} ;\citep[44 =][]{mcconnell12} ; \citep[45 =][]{HRix} ; \citep[46 =][]{Magorrian} ; \citep[47 =][]{GT09}.\\
\\
Notes on individual galaxies:\\
\\
$^a$  The literature contains a large number of estimates for the velocity dispersion of our Galaxy's bulge, using different kinematic tracers at different radii.  We use the radially averaged measurement of $\sigma = 103 \pm 20 \kms$ from \citet{Tremaine02}. \\
\\
$^b$  We have re-computed $\sigma$ for 12 galaxies, considering kinematic data between $\rinf$ and $\reff$.  The corresponding values of $\sigma$ are listed here.  Table~\ref{tab:rinf} also lists the values of $\sigma$ using data from 0 to $\reff$. \\
\\
$^c$  Maser-based black hole masses for several galaxies are presented in \citet{Greene10} and \citet{Kuo11}.  We use the velocity dispersions presented in \citet{Greene10}.  For consistency with the rest of our sample, we use the black hole masses from \citet{Kuo11}, which agree with the values in \citet{Greene10} but do not include distance uncertainties in the overall uncertainty for $\mbh$.  \citet{braatz10} provide an updated distance and black hole mass for UGC 3789, which are consistent with the values we adopt from \citet{Kuo11}. \\
\\
$^d$  Following G09, our sample includes two distinct measurements for NGC 1399.  We weight each of these measurements by 50\% when performing fits to the black hole scaling relations.  \\
\\
$^e$  The literature contains two inconsistent estimates of the $V$-band luminosity of NGC 3607: $M_V = -21.62$ in G09, and $M_V = -19.88$ in \citet{Lauer07}.   \\
\\
$^f$  The literature contains two inconsistent estimates of the $V$-band bulge luminosity of NGC 4564: $M_V = -19.60$ in G09, and $M_V = -20.26$ in \citet{Lauer07}.  \\
\\
$^g$  Bulge luminosities for NGC 4736 and NGC 4826 were included in the sample of \citet{mcconnell11b} and their fit to the $\mbh-L$ relation.  These luminosities corresponded to pseudobulges identified in \citet{KBC11}, and we have not included them in our present fits.\\
\\
$^h$  The stellar dynamical measurement of $\mbh$ in NGC 5128 by \citet{Capp09} is fully consistent with the molecular gas measurement $\mbh = 5.3^{+0.6}_{-0.4} \msun$ by \citet{Neumayer07}.\\
\\
$^w$ We have updated the distances to 29 galaxies using surface brightness fluctuation measurements from \citet{Tonry01}, with the corrections suggested by \citet{Blakeslee10}.\\
\\
$^x$ We have updated the distances to 12 galaxies using surface brightness fluctuation measurements from \citet{Blakeslee09}, which are based on data from ACS on the \textit{Hubble Space Telescope}.\\
\\
$^y$ We have adopted a distance of 0.73 Mpc to M31, based on Cepheid variable measurements by \citet{Vilardell07}.  We assume that M31 and M32 lie at the same distance.\\
\\
$^z$  For NGC 4342, we have adopted the distance of 23 Mpc by \citet{Bogdan12b}.

\end{table*}
\clearpage

\end{document}